\documentclass{jfm}

\usepackage{graphicx}
\usepackage{natbib}
\usepackage{epstopdf}
\ifCUPmtlplainloaded \else
  \checkfont{eurm10}
  \iffontfound
    \IfFileExists{upmath.sty}
      {\typeout{^^JFound AMS Euler Roman fonts on the system,
                   using the 'upmath' package.^^J}%
       \usepackage{upmath}}
      {\typeout{^^JFound AMS Euler Roman fonts on the system, but you
                   dont seem to have the}%
       \typeout{'upmath' package installed. JFM.cls can take advantage
                 of these fonts,^^Jif you use 'upmath' package.^^J}%
      }
  \else
  \fi
\fi

\ifCUPmtlplainloaded \else
  \checkfont{msam10}
  \iffontfound
    \IfFileExists{amssymb.sty}
      {\typeout{^^JFound AMS Symbol fonts on the system, using the
                'amssymb' package.^^J}%
       \usepackage{amssymb}%
         \let\leq=\leqslant
         
      }{}
  \fi
\fi

\ifCUPmtlplainloaded \else
  \IfFileExists{amsbsy.sty}
    {\typeout{^^JFound the 'amsbsy' package on the system, using it.^^J}%
     \usepackage{amsbsy}}
    {}
\fi

\newsavebox{\astrutbox}
\sbox{\astrutbox}{\rule[-5pt]{0pt}{20pt}}

\usepackage{subfig}
\usepackage{float}
\usepackage[usenames]{color}
\usepackage{natbib}

\title{Nonlinear waves on the surface of a fluid covered by an elastic sheet}
\author[Luc Deike, Jean-Claude Bacri and Eric Falcon]
{Luc Deike%
  \thanks{Email address for correspondence: luc.deike@univ-paris-diderot.fr},\ns
Jean-Claude Bacri and Eric Falcon}

\affiliation{Univ Paris Diderot, Sorbonne Paris Cit\'e, MSC, UMR 7057 CNRS, F-75 013 Paris, France, EU}

\date{\today}

\begin{document}

\maketitle

\begin{abstract} 
We experimentally study linear and nonlinear waves on the surface of a fluid covered by an elastic sheet where both tension and flexural waves take place. An optical method is used to obtain the full space-time wave field, and the dispersion relation of waves. When the forcing is increased, a significant nonlinear shift of the dispersion relation is observed. We show that this shift is due to an additional tension of the sheet induced by the transverse motion of a fundamental mode of the sheet. When the system is subjected to a random noise forcing at large scale, a regime of hydro-elastic wave turbulence is observed with a power-law spectrum of the scale in disagreement with the wave turbulence prediction. We show that the separation between relevant time scales is well satisfied at each scale of the turbulent cascade as expected theoretically. The wave field anisotropy, and finite size effects are also quantified and are not at the origin of the discrepancy. Finally, the dissipation is found to occur at all scales of the cascade contrary to the theoretical hypothesis, and could thus explain this disagreement.
\end{abstract}

\section{Introduction}
Wave turbulence theory, or weak turbulence, concerns the dynamical and statistical properties of a set of nonlinear interacting waves (\cite{ZakharovBook, NazarenkoBook, NewellReview}). This theory has been developed at the end of the 60's, and has been applied to various domains: surface waves in oceanography, Rossby waves in the atmosphere, spin waves in magnetic materials, Kelvin waves in superfluid turbulence, and nonlinear optics. It assumes strong hypotheses such as those addressing weakly nonlinear, isotropic and homogeneous random waves in an infinite size system with scale separation between injection and dissipation of energy. It notably predicts analytical solutions for the spectrum of a weakly nonlinear wave field at equilibrium or in a stationary out-of equilibrium regime. So far, few well controlled experiments on wave turbulence have been designed to specifically test the theory and its limitations. They mainly concern capillary-gravity waves on the surface of a fluid (\cite{Putterman1996,Falcon07a,KolmakovBook,Xia10,Denissenko07,Falcon2010,Cobelli2011}), elastic waves on a metallic plate (\cite{Mordant2008,Boudaoud08}), and nonlinear optics (\cite{ResidoriReview}).  Experimental studies of wave turbulence on original systems are thus of great interest to notably test the domain of validity of the theory in experiments.

Here, we present a well controlled experiment to study nonlinear interacting waves on the surface of an elastic sheet covering the surface of a fluid. Two types of hydroelastic waves take place: tension waves and flexural (or bending) waves that result from the coupling of the elastic sheet response with the underneath fluid one. The main motivation is to study wave turbulence in this new system that enables to tune the dispersion relation of waves by controlling the elastic sheet tension applied.

It is noteworthy that the response of a thin elastic sheet covering a fluid to a dynamical perturbation usually generates hydroelastic surface waves occurring in various domains: bio-medical applications such as heart valves (\cite{Grotberg2004}), flapping flags (\cite{ZhangReview}) and industrial applications like very large floating structures (\cite{WatanabeReview}). In oceanography, flexural-gravity waves are known to propagate on the surface of lakes or oceans covered by ice and are involved in various ice floe phenomena (\cite{Takizawa1985, SquireNature, SquireBook, SquireReview2007, Dias}). Such waves are governed by the same wave equation than the one presented here. However, this analogy is quite limited since the free floating boundary condition in the in-situ case is not respected here, as well the orders of magnitude leading to the predominance of in-situ flexural-gravity waves instead of flexural-tension waves here.

The paper is organized as follows. The theoretical background is presented in $\S$ \ref{theory}. The linear equations and the boundary conditions of an elastic sheet covering a fluid are recalled. The relation between an applied pressure to the sheet and the induced static tension is then given. The involved nonlinearities in the system are discussed as well as their implications on wave interactions processes. Dimensional analysis of wave turbulence is then presented for the case of tension and flexural waves on the surface of a fluid covered by an elastic sheet. In $\S$ \ref{setup}, the experimental setup is presented. The optical techniques to measure the wave field are introduced and the tuning of the static tension of the sheet by an applied hydrostatic pressure. Results are presented in $\S$ \ref{dispersion} and \ref{wt}. In $\S$ \ref{dispersion}, we have observed tension and bending waves on two decades in frequency in good agreement with the theoretical dispersion relation of linear waves. When the wave amplitude is increased, a shift of the dispersion relation occurs. We show that this shift is due to the dynamics of slow nonlinear waves (fundamental eigenmode of the sheet) that induces an additional quasi-static tension to the sheet. The observation of the wave turbulence regime is then presented $\S$ \ref{wt}. The spectrum of the wave transverse velocity is found to scale as a power law of both the frequency and wave number, both power law exponents being in disagreement with the ones predicted by wave turbulence theory. We then show that wave anisotropy exists but is not at the origin of this discrepancy. The dissipation time of the waves, the nonlinear interaction time and the linear time of wave propagation have been measured at each scale, and show that the time scale separation hypothesis of wave turbulence theory is fulfilled. On the other hand, dissipation is found to occur at all scales contrary to the theoretical hypothesis, and could thus explain this disagreement for the spectrum scaling. The conclusions are drawn in $\S$ \ref{conclusion}.

\section{Theoretical background \label{theory}}
\subsection{Linear dispersion relation \label{linear}}
Assume a floating elastic sheet subjected to an uniform and isotropic tension $T$. Properties of the sheet are: density $\rho_{e}$, Young modulus $E$, Poisson modulus $\nu$, and thickness $h$. The fluid density is $\rho$ and its depth beneath the sheet at rest $H$. The momentum equation of the thin elastic sheet is then given by (\cite{Landau,Sneyd1985,Sneyd1987})
\begin{equation}
D\nabla^{4}\eta-T\nabla^{2}\eta+\rho_{e}h\frac{\partial^2 \eta}{\partial t^2}=p,
\label{pfd}
\end{equation}
with $\eta$ the vertical sheet deformation, $p$ the pressure due to the liquid on the elastic sheet, $D\equiv \frac{Eh^{3}}{12(1-\nu^{2})}$ the bending modulus of the elastic sheet, $\nabla^2 \equiv \partial^2/\partial x^2 + \partial^2/\partial y^2$, and $\nabla^4\equiv \partial^4/\partial x^4 + \partial^4/\partial y^4 + 2\partial^4/\partial x^2\partial y^2$ the Laplacian and bi-Laplacian operators. Considering an irrotational flow of velocity potential $\phi(x,y,z,t)$, the pressure equation on the surface $z=\eta$ is given by
$p=-\rho\frac{\partial \phi}{\partial t}_{z=\eta}-\rho g\eta$, with $g$ the acceleration of gravity.
For a plane wave solution, using the kinematic boundary condition on the sheet surface $\frac{\partial \eta}{\partial t}=\frac{\partial \phi}{\partial z}_{z=\eta}=\tanh{kH}(\phi)_{z=\eta}$, and assuming negligible sheet inertia ($\rho_e kh/\rho\ll 1$) and infinite depth ($kH\gg1$), the linear dispersion relation reads (\cite{Sneyd1987})
\begin{equation}
\omega^{2}=gk+\frac{T}{\rho}k^{3}+\frac{D}{\rho}k^{5}.
\label{LDR}
\end{equation}
It involves three terms: a gravity one, and two elastic terms. The second term of the right hand member of Eq.\ (\ref{LDR}) is a tension term structurally analogous of a capillary term (\cite{Lamb}), whereas the third term corresponds to bending. Note that the dispersion relation for pure elastic waves on a plate is $\omega^2\sim Tk^2+ Dk^4$ (\cite{Landau}), Eq.\ (\ref{LDR}) coming from the coupling of the sheet elasticity with the underneath liquid. The crossover between the various wave regimes in Eq.\ (\ref{LDR}) can be evaluated, for typical values of the applied tension $T\approx4$ N/m and of the bending modulus $D\approx 5\ 10^{-6}$ Nm of the latex sheet used here (see $\S$\ref{setup}). Balancing the first and second terms of the right hand member of Eq.\ (\ref{LDR}) gives the transition between gravity and tension waves, $\lambda_{gT}= 2\pi \sqrt{T/(g\rho)} \simeq 10$ cm. Balancing the second and third terms of the right hand member of Eq.\ (\ref{LDR}) gives the transition between bending and tension waves, $\lambda_{TD}= 2\pi \sqrt{D/T} \simeq 1$ cm. Note that for ice floes, the bending elastic term prevails over the tension one, and the order of magnitude of the flexural-gravity transition is $\lambda_{gD}= 2\pi [D^{\star}/(g\rho)]^{1/4} \simeq 100$ m for a typical ice bending modulus $D^{\star}\approx 10^9$ Nm (\cite{Sneyd1985, Sneyd1987}).

\subsection{Boundary conditions and sheet eigenmodes \label{eigen}}
Let us consider an elastic sheet clamped on the top of a vertical circular vessel of radius $R$. The boundary condition is thus $\eta(R)=0$ (and consequently $\partial \eta(R)/\partial t=0$). In such a circular geometry, the sheet eigenmodes are given by the zeros of the Bessel functions: $J_n(kR)=0$, with $J_n$ a Bessel function of the first kind, $n$ an integer and $k$ the eigenmode wave number (\cite{Morse}). The first symmetric and antisymmetric eigenvalues are $J_0(kR)=0$ leading to $k_{0,1}R=2.405$, and $k_{0,2}R=5.520$, and $J_1(kR)=0$ leading to $k_{1,1}R=3.832$, and $k_{1,2}R=7.016$. The corresponding natural frequencies are then obtained using the dispersion relation of Eq. \ref{LDR}. Note that in the case of an ice floes, the ice boundary is free to move and free edges boundary conditions has to be considered ($\partial \eta(R)/\partial t=0$ and vanishing of the bending moment, see \cite{SquireBook}).

\subsection{Scaling of tension with pressure: Static case \label{TvsP}}
An external pressure $P_s$ applied on an elastic sheet clamped on the top of a circular vessel generates a static tension $T_s$ of the sheet. $T_s$ is analytically computed as function of $P_s$ as follows. When $P_s$ is applied, the sheet deforms, and for large enough deformation (maximum deflexion $\eta_m\gg h$), the tension term prevails over the bending one in Eq.\ (\ref{pfd}) (\cite{Landau}). Under the assumption of an homogeneous and isotropic stress tensor, the static force equilibrium can be written as $T_s\nabla^2\eta=P_s$ (\cite{Landau}). Using the circular boundary condition $\eta(R)=0$, the deformed solution for a sheet of radius $R$ is a parabolic surface $\eta(r)$ (\cite{Landau})
\begin{equation}
\eta(r)=\frac{P_s}{4T_s}(R^2-r^2).
\label{shape}
\end{equation}
The maximum deflexion of the parabola thus reads: 
\begin{equation}
\eta_m \equiv \eta(0)=P_sR^2/(4T_s).
\label{maxdeflex}
\end{equation}
Under the above hypotheses and using (\cite{Landau}) the dependence between the $P_s$ and $T_s$ can be calculated: 
\begin{equation}
T_s=\left[\frac{Eh}{32(1-\nu)}P_s^2R^2\right]^{1/3}.
\label{stth}  
\end{equation}

\subsection{Nonlinear resonant interactions \label{inter}}
Nonlinearities involved in the system can be introduced either in the equations of the elastic plate, or in the ones of the fluid, or in the boundary condition between the fluid and the plate. A complete discussion of the corresponding equations can be found in \cite{Peake2001,Peake2006}. Although the equation for a single elastic plate involves cubic nonlinearities, the pressure term and the boundary condition between the fluid and the plate involves quadratic nonlinearities. Indeed, the pressure term reads
\begin{equation}
p(x,y,t)=\rho g \eta + \frac{\partial \phi}{\partial t}+ \rho \frac{v^2}{2},
\label{eqpnl}
\end{equation}
with $\eta(x,y,t)$ the wave height, and $v(x,y,t)=\frac{\partial \eta}{\partial t}$ the transverse wave velocity. The importance of the nonlinear term in the pressure will be shown experimentally in $\S$ \ref{dispersion}. Thus, the coupling between the fluid and the elastic plate involves a quadratic nonlinearity of the pressure term. 

When considering resonant wave interactions, 3-wave interactions occur when quadratic nonlinearities are present in the system while 4-wave interactions have to be considered in the case of cubic nonlinearities (\cite{ZakharovBook,NazarenkoBook}). The dynamics of wave interactions is dominated by the lowest nonlinear order (\cite{ZakharovBook, NazarenkoBook}). Thus, when 3-wave and 4-wave interactions occur as in our case, 4-wave interactions are generally neglected. Thus, only 3-wave interactions should be considered here. 3-wave interactions have also to be compatible with the dispersion relation. In our case, the dispersion relation of tensional and bending waves is of decay type, i.e. $\omega\sim k^{\mu}$ with $\mu>1$ (as for capillary waves), and thus fulfills the 3-wave resonant conditions on the frequency ($\omega_1\pm\omega_2\pm\omega_3=0$) and on the wave numbers ($\mathbf{k}_1\pm\mathbf{k}_2\pm\mathbf{k}_3=0$). A kinetic equation assuming 3-wave interactions has been proposed in the case of floating ice sheet on water \cite{Marchenko}. Note also that in the case of pure flexural wave turbulence on an elastic plate (without water), 4-wave interactions are considered since the nonlinearity of the plate are cubic,  (\cite{During2006}). For pure gravity waves, nonlinearities are quadratic but 3-wave interactions are not possible since the resonant conditions are not satisfied due to the geometry of the dispersion relation  ($\mu<1$), and only 4-wave interactions have to be considered (\cite{McGoldrick1966} and references therein).

\subsection{Dimensional analysis in wave turbulence}
One of the main results of wave turbulence theory is the existence of out-of-equilibrium stationary solutions of the kinetic-like equation for the wave action spectrum $n_k=E_k/\omega$, with $E_k$ the wave energy spectrum in the Fourier space. The kinetic equation generally reads (\cite{ZakharovBook, NazarenkoBook,NewellReview}):
\begin{equation}
\frac{\partial n_k}{\partial t}=St(n_k)+F_k-\Gamma_k n_k
\label{kinetic}
\end{equation}
with $St(n_k)$ is the collision integral that depends on the physical properties of the propagation medium and on the type of nonlinear wave interaction, $F_k$ is the forcing term, and $\Gamma_k$ the dissipative rate. Energy conservation leads to a stationary solution in the form of a direct energy cascade from the injection scale $k_i$ to the dissipation scale $k_d$, the cascade inertial range being defined by the scale separation hypothesis $k_i\ll k\ll k_d$ in a similar way to hydrodynamic turbulence. The other main hypothesis used to derive the wave turbulence spectrum is an infinite medium, local interactions, an homogeneous and isotropic wave field, weak nonlinearities and the following time scale separation: $\tau \ll \tau_{nl} \ll \tau_d$, with $\tau=1/\omega$ the linear propagation time, $\tau_{nl}$ the nonlinear wave interaction time and $\tau_d$ the dissipation time (\cite{NewellReview,ZakharovBook,NazarenkoBook}).

Under the previous hypothesis, dimensional analysis can be performed in wave turbulence as presented by \cite{ConnaughtonDA} in order to determine stationary solutions of the kinetic equation. For a given dispersion relation $\omega=ck^{\mu}$, the wave energy spectrum of a set of nonlinear waves is supposed to be a self-similar function of the scale $k$, the flux of the conserved quantity and the constant $c$. Here, direct cascade of energy is considered and the flux through the scales is the energy flux per mass density, $\epsilon$ (dimension $[L^3T^{-3}]$). Experimentally, the quantity of interest here is the power spectrum of wave vertical velocity, $S_v(k)$ (dimension $[L^3T^{-2}]$) that is assumed to read 
\begin{equation}
S_{v}(k) \sim \epsilon^{1/(N-1)}c^{\beta}k^{\alpha},
\end{equation}
with $N$ the number of interacting waves in the leading process. The dependency $S_{v}(k) \sim \epsilon^{1/(N-1)}$ is given by wave turbulence theory (\cite{ConnaughtonDA,ZakharovBook,NazarenkoBook}). As explained above in $\S$ \ref{inter}, $N=3$ has to be considered in our case. For tensional waves in deep water, one has $\omega^2=\frac{T}{\rho}k^{3}$, and one gets dimensionally 
\begin{equation}
S^{t}_{v}(k) \sim \epsilon^{1/2}\left(\frac{T}{\rho}\right)^{1/4}k^{-3/4},
\label{DAtension}
\end{equation}
which is analog to the case of capillary waves. For bending waves in deep water, one has $\omega^2=\frac{D}{\rho}k^{5}$, we get dimensionally 
\begin{equation}
S^{b}_{v}(k) \sim \epsilon^{1/2}\left(\frac{D}{\rho}\right)^{1/4}k^{-1/4}.
\label{DAflex}
\end{equation}
Similarly, one obtains the frequency spectra of the wave height (dimension $[L^2T]$)
\begin{equation}
S^{t}_{\eta}(f) \sim \epsilon^{1/2}\left(\frac{T}{\rho}\right)^{1/6}f^{-17/6}, \ {\rm and}\ S^{b}_{\eta}(f) \sim \epsilon^{1/2}\left(\frac{D}{\rho}\right)^{1/10}f^{-27/10}. 
\label{DAtime}
\end{equation}
The scaling of the nonlinear interaction time is determined by balancing the left-hand side and the collision integral of Eq. (\ref{kinetic}), $1/\tau_{nl} \sim St(n_k)/n_k$ (\cite{ConnaughtonDA,Newell2001}). Using Eq. (\ref{DAtension}) and the dimensional structure of the collision integral for 3-wave interactions for tensional waves, lead to 
\begin{equation}
1/\tau_{nl} \sim\epsilon^{1/2}\left(\frac{T}{\rho}\right)^{-1/4}k^{3/4},
\label{tnl_dim}
\end{equation}
while for bending waves:
\begin{equation}
1/\tau_{nl} \sim\epsilon^{1/2}\left(\frac{D}{\rho}\right)^{-1/4}k^{1/4}.
\label{tnl_dim}
\end{equation}

\section{Experimental setup\label{setup}}
\begin{figure}
\begin{center}
\includegraphics[scale=0.45]{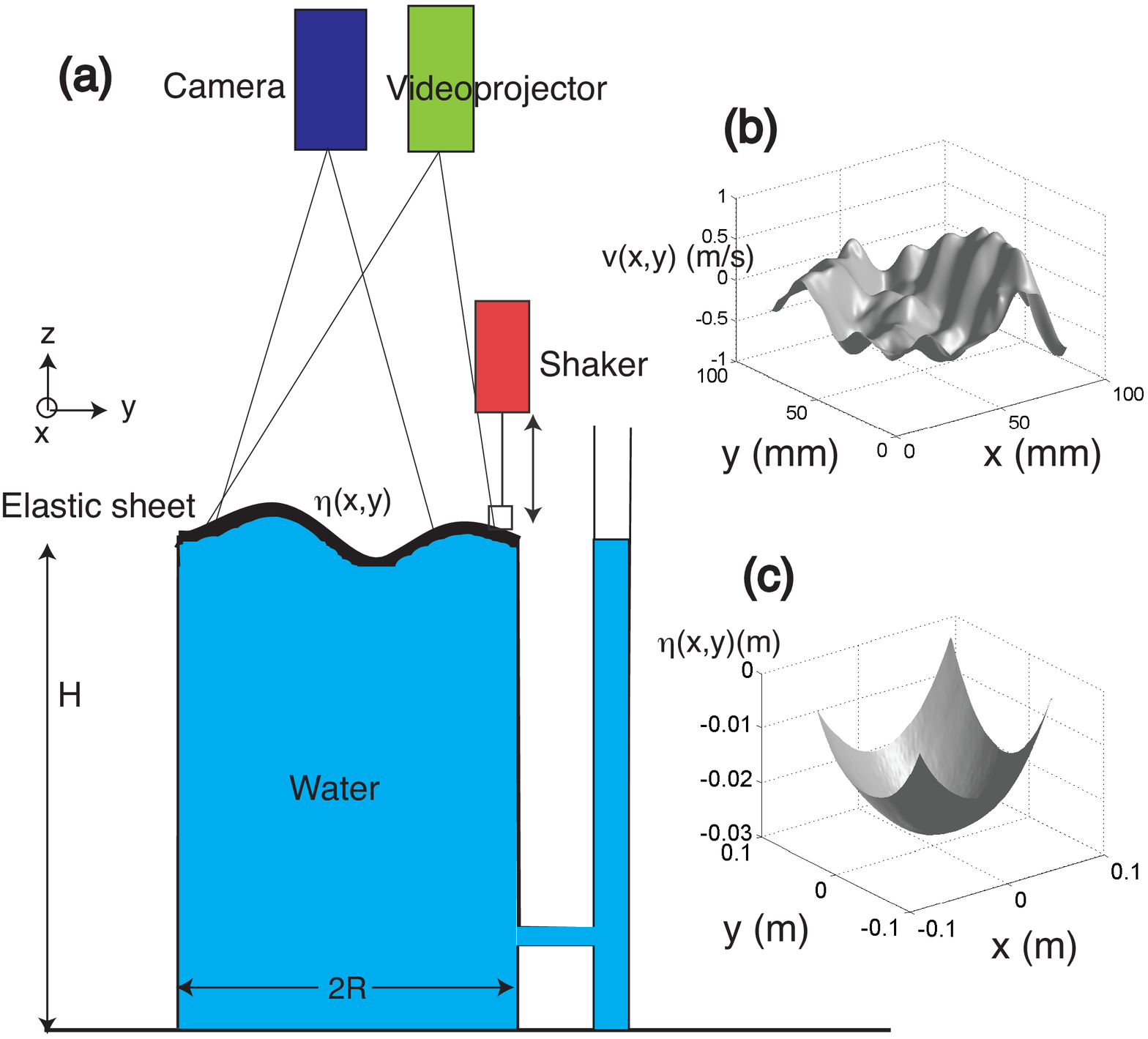}
\includegraphics[scale=0.4]{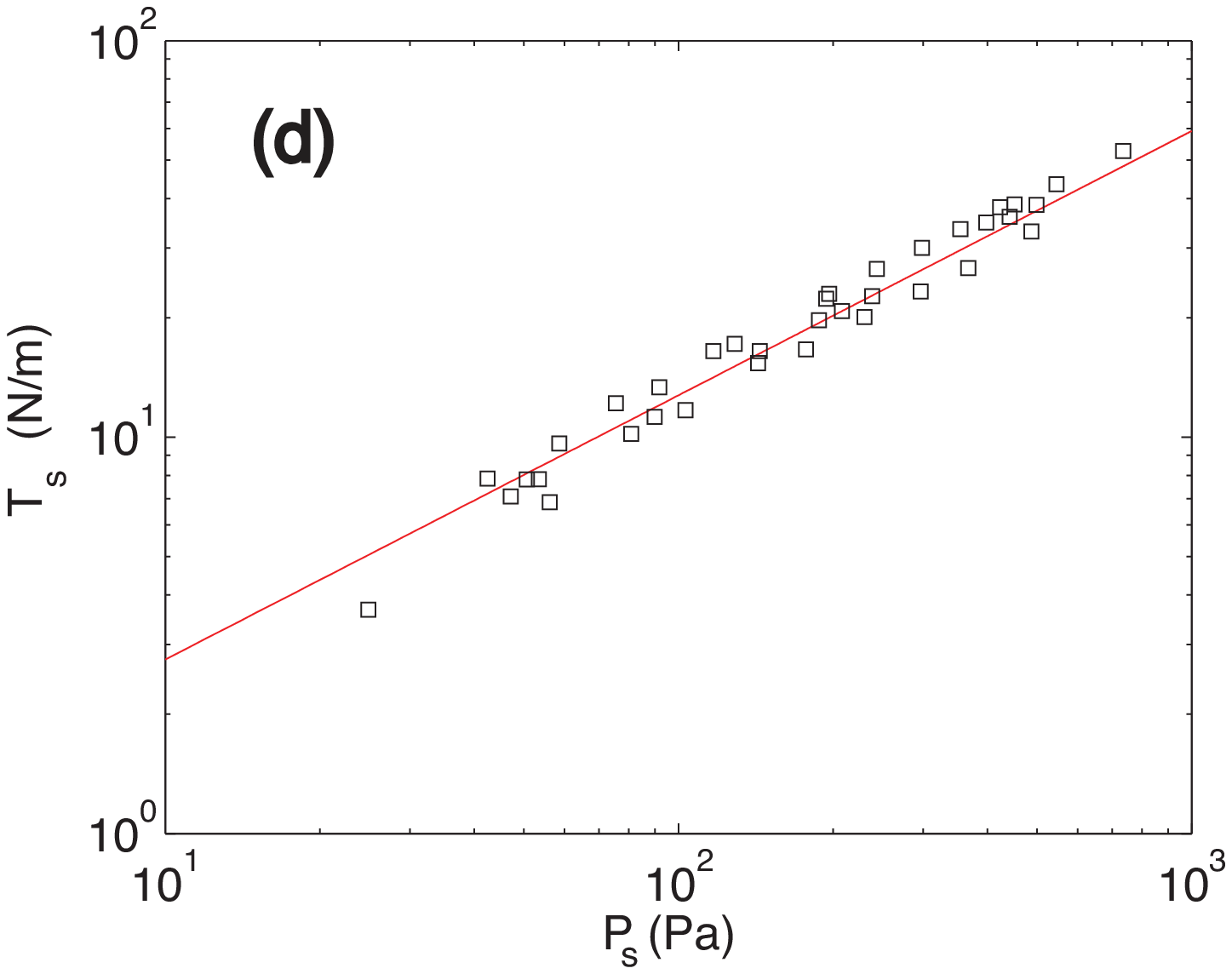}
\caption{(a) Experimental setup. (b) Typical wave transverse velocity $v(x,y)$ reconstructed by Fourier transform profilometry. (c) Reconstructed height profile of the sheet for $P_s=-130$ Pa showing a parabolic static profile. (d) Static tension $T_s$ as a function of the applied pressure $|P_s|$. Solid line corresponds to Eq. (\ref{stth}), $T_s=\left[\frac{Eh}{32(1-\nu)}P_s^2R^2\right]^{1/3}$.}
\label{exp}
\end{center}
\end{figure}
The experimental setup is shown in Fig. \ref{exp}a. It consists of a cylindrical vessel of radius $R=10$ cm (or 20 cm) filled with water up to a height $H=43$ cm (or 20 cm, respectively). An elastic latex sheet is stuck on the top circular side of the container. We carefully checked that no air bubbles are trapped between the sheet and the water. Measured physical properties of the latex sheet are: thickness $h=0.35$ mm (or 0.5 mm), Young modulus $E=1.05\ 10^6$ N/m$^2$ (or $1.5 \ 10^6$ N/m$^2$, respectively), and Poisson modulus $\nu\approx0.5$ (industrial latex were provided by Carrat and Eurocatsuits). Waves on the sheet are generated by the vertical motion of a rectangular wave maker ($75 \times 10$ mm, or $186 \times 10$ mm) driven by an electromagnetic shaker (LDS V201 or LDS V406) as shown in Fig.\ \ref{exp}a. The shaker is driven by a sinusoidal forcing at frequency $f_p$ (with $f_p\in[10:50]$ Hz) or by a random noise forcing band-pass filtered around $f_p$  within a frequency bandwidth $f_p \pm \Delta f$ (with $f_p\in[8:15]$ Hz and $\Delta f=1.5$ Hz). A stationary wave field state is reached after a few seconds and measurements are done during the steady state. The full 3D space-time wave field is measured using a fast Fourier profilometry technique (\cite{Cobelli2009}), recently used on elastic waves on a metallic plate (\cite{Mordantepj}) or for gravity-capillary waves on a fluid surface (\cite{Falcon2010,Cobelli2011}). Fringes with interfranges of 1 mm are projected on the sheet surface by a high resolution video-projector (Epson TW3000), and the space-time evolution of the fringe deformations enables to reconstruct the  velocity normal to the free surface $v(x,y,t)$ with a fast camera (Phantom V9) recording at 1000 fps during $\mathcal{T}=4$ s. The size of the recorded images, centered in the middle of the sheet, is $10 \times 8$ cm$^2$ for the small vessel, and $25 \times 20$ cm$^2$ for the large one. A typical reconstructed pattern of $v(x,y,t)$ is shown at a given time in Fig. \ref{exp}b. From the movie of $v(x,y,t)$, one computes the power spectrum density of transverse velocity $S_v(k_x,k_y,f)$ from multidimensional Fourier transform. By integrating $S_v(\mathbf{k},f)$ over all directions of the wave vector $\mathbf{k}$, we also obtain $S_v(k=||\mathbf{k}||,f)$, with $k\equiv \sqrt{k_x^2+k_y^2}$ the wave number. The transverse velocity of the wave at a fixed location has been also directly recorded with a Doppler velocimeter (Polytech OPV 506) for long times, $\mathcal{T}=300$ s, to measure the slow response of the system. The wave maker velocity $V(t)$ is measured by an accelerometer (BK 4393) fixed on the wave maker and the force $F(t)$ applied by the wave maker on the sheet is measured by a piezoelectric sensor (FGP 10 daN). The mean power $I$ injected within the system is then evaluated by $I = \frac{1}{\mathcal{T}} \int_0^{\mathcal{T}} F (t)V (t)dt$ (\cite{FalconReview}).

As shown in Fig. \ref{exp}a, the vessel is connected through a small pipe to a vertical tube that enables the control of the hydrostatic pressure $P_s$ imposed on the elastic sheet by just adding or removing an amount of water from the tube. When the water inside the vertical tube is at the same height than the sheet, $P_s=0$. By filling (or draining) the tube of water up to a height $\Delta H$ of water, an external positive (or negative) pressure is applied since the sheet is stuck on the top of the vessel. As shown in Fig.\ \ref{exp}c, the elastic sheet has a parabolic shape well described by Eq. (\ref{shape}), and its maximum deflexion $\eta_m$ can be then measured. The hydrostatic pressure is then estimated by $P_s=\rho g (\Delta H - \eta_m)$, since the pressure between the top and the bottom of the parabola is negligible. Note that this condition is needed to consider a constant pressure over the whole elastic sheet. For various applied $P_s$, we measure $\eta_m$, and experimentally deduce the corresponding static tension $T_s$ on the sheet from Eq.\ (\ref{maxdeflex}). As shown in Fig. \ref{exp}d, $T_s$ is found to evolve with $P_s$ in good agreement with Eq. (\ref{stth}) with no fitting parameter during this comparison. Note that the depression or overpressure cases are similar since the weight of the sheet is negligible for our range of $P_s$. Thus the absolute value of $P_s$ will be given in the following. The initial static tension $T_s$ on the sheet is thus well controlled by the applied static pressure $P_s$. This allows us to study the linear and nonlinear properties of waves on a floating elastic sheet with a tunable tension. 

The pressure only due to the weight of the elastic sheet with no water beneath is $P_s^0=\rho_e g h$. One experimentally gets from Eq.\ (\ref{maxdeflex}) the static tension $T_s^0$ of the elastic sheet due to its sticking on the vessel without water. One finds $T_s^0\simeq 2$ to 4 N/m whatever the elastic sheets and container sizes used. This value will correspond to the minimal value of static tension reached when water is added and no hydrostatic pressure applied. Thus, one considers $P_s=P_s^0$ and $T_s=T_s^0$ in this case. As pointed out in $\S$ \ref{linear}, for $T_s^0\approx4$ N/m, the transition between gravity and tensional waves is $\lambda_{gT} \simeq 10$ cm, of the order of the vessel size and the observation window. Thus, only elastic waves will be observed here. Moreover, bending waves will occur for wavelengths smaller than $\lambda_{TD}\simeq 1$ cm. In a prestressed case, $T_s$ is increased and these transitions change as $\lambda_{gT}\sim \sqrt{T_s}$ and $\lambda_{TD}\sim 1/\sqrt{T_s}$ (see $\S$ \ref{linear}). Thus, almost only tensional waves will be observed in the case of a large initial tension: for instance, if $T_s=10$ N/m, $\lambda_{gT} \simeq 20$ cm and $\lambda_{TD}\simeq 0.4$ cm.

\section{Dispersion relation of nonlinear waves\label{dispersion}}
Here, we characterize experimentally the dispersion relation of waves when the system is subjected to either a monochromatic or a filtered random noise forcing.

\begin{figure}
\begin{center}
\includegraphics[scale=0.45]{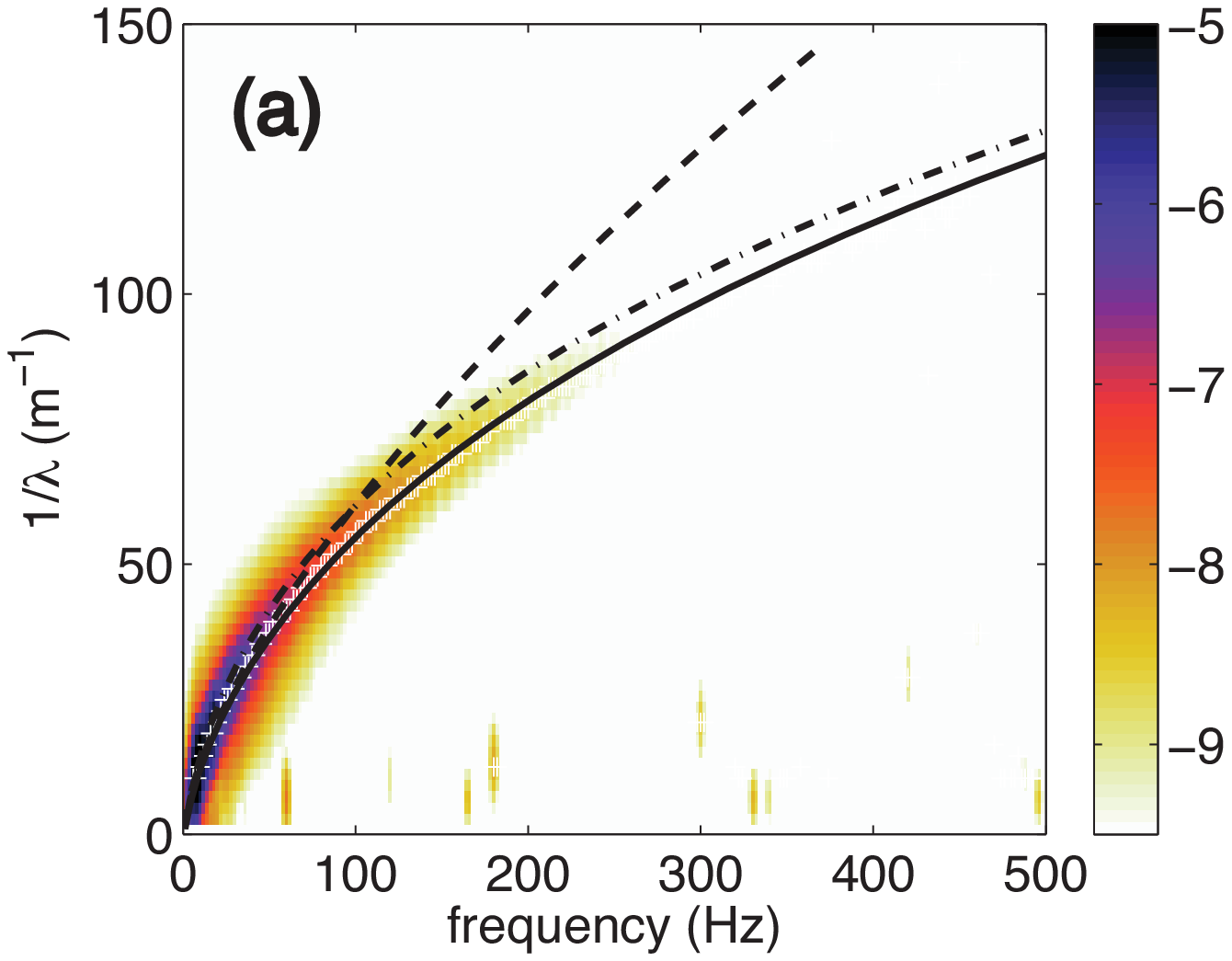}
\includegraphics[scale=0.45]{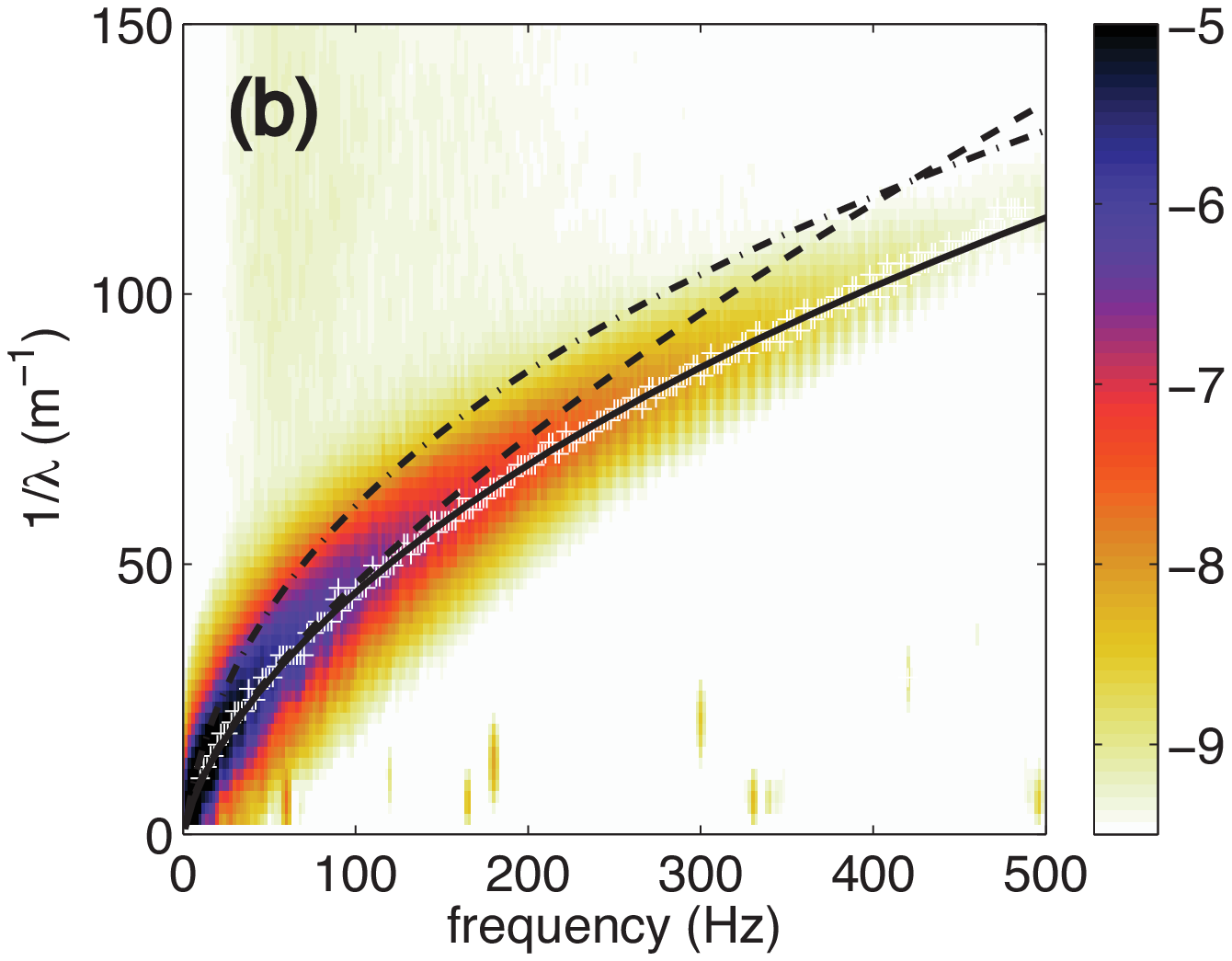}
\end{center}
\caption{Space-time spectrum $S_v(k,f)$ of the transverse velocity of waves for: (a) moderate ($p=12$ Pa) and (b) strong ($p=33$ Pa) forcing amplitudes. Random forcing bandwidth: $8.5\pm 1.5$ Hz (a). Colors are log scaled. $P_s=0$. ($+$) correspond to the local maxima of $S_v(k,f)$ for each frequency. Dashed white lines are $\omega^2=\frac{T}{\rho}k^3$. Solid lines are $\omega^2=\frac{D}{\rho}k^5+\frac{T}{\rho}k^3$. Dash-dot lines are $\omega^2=\frac{D}{\rho}k^5+\frac{T_s^0}{\rho}k^3$. Fitting parameter: $T=$ (a) 7 and (b) 16 N/m. $f^*=20$ Hz and $1/\lambda^*=20$ m$^{-1}$ from Eq.\ (\ref{LDR}).
\label{fig02}}
\end{figure}

\subsection{Dispersion relation}
The space-time power spectrum $S_v(k,f)$ of the transverse velocity of waves is shown Fig.~\ref{fig02} for two forcing amplitudes, and for $P_s=0$. In both cases, the wave energy injected at low frequencies is transferred through the scales towards high frequencies by nonlinear interaction between waves. The wave energy is mainly localized on a single curve in the ($\omega\equiv 2\pi f$,$k\equiv2\pi/\lambda$) space that corresponds to the non-linear wave dispersion relation. Both theoretical dispersion relations of pure tension waves, $\omega^2=\frac{T}{\rho}k^3$, and of tension and bending waves, $\omega^2=\frac{D}{\rho}k^5+\frac{T}{\rho}k^3$ are plotted in Fig. \ref{fig02} with $T$ the single fitting parameter, as well as the dispersion relation for the tension equal to the static tension $T=T_s^0$. When both elastic terms are taken into account, the theoretical dispersion relation well describes the experimental results over the whole frequency range. Tension waves occur mainly at low frequencies whereas bending waves occur at higher frequencies as expected. When the forcing amplitude is increased (see Fig.\ \ref{fig02}b), wave interactions still redistribute the wave energy on the dispersion relation up to higher frequencies, but with two main differences. First, the width of the dispersion relation is broader in $f$ and $k$ directions. Second, a shift of the dispersion relation is observed, the fitting value of $T$ being more than doubled when the forcing is increased by a factor 3. Thus, the observed nonlinear dispersion relation follows the linear predicted one with a tension that depends on the forcing amplitude.

\subsection{Shift of the dispersion relation}
\label{shift}
\begin{figure}
\begin{center}
\includegraphics[scale=0.42]{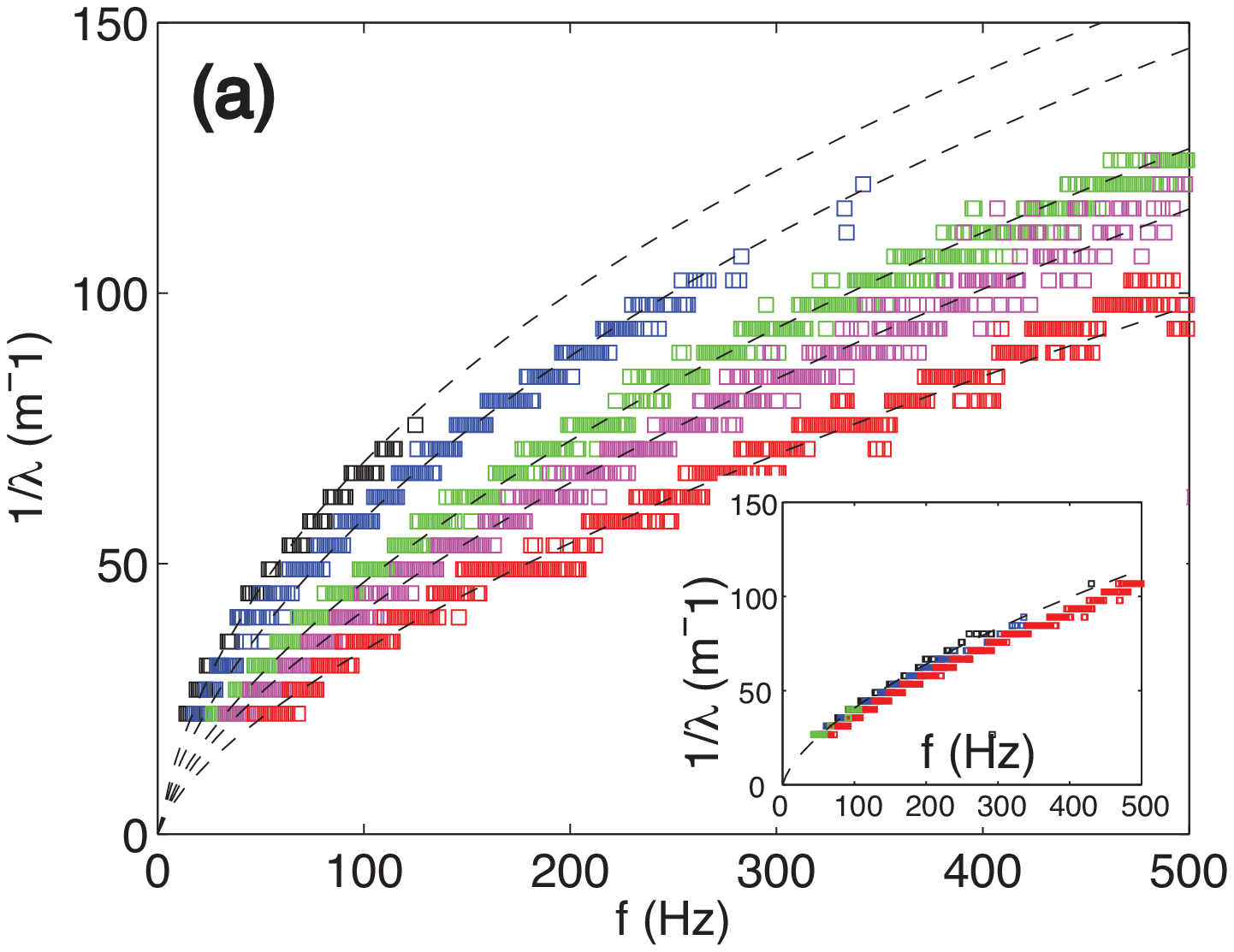}
\includegraphics[scale=0.45]{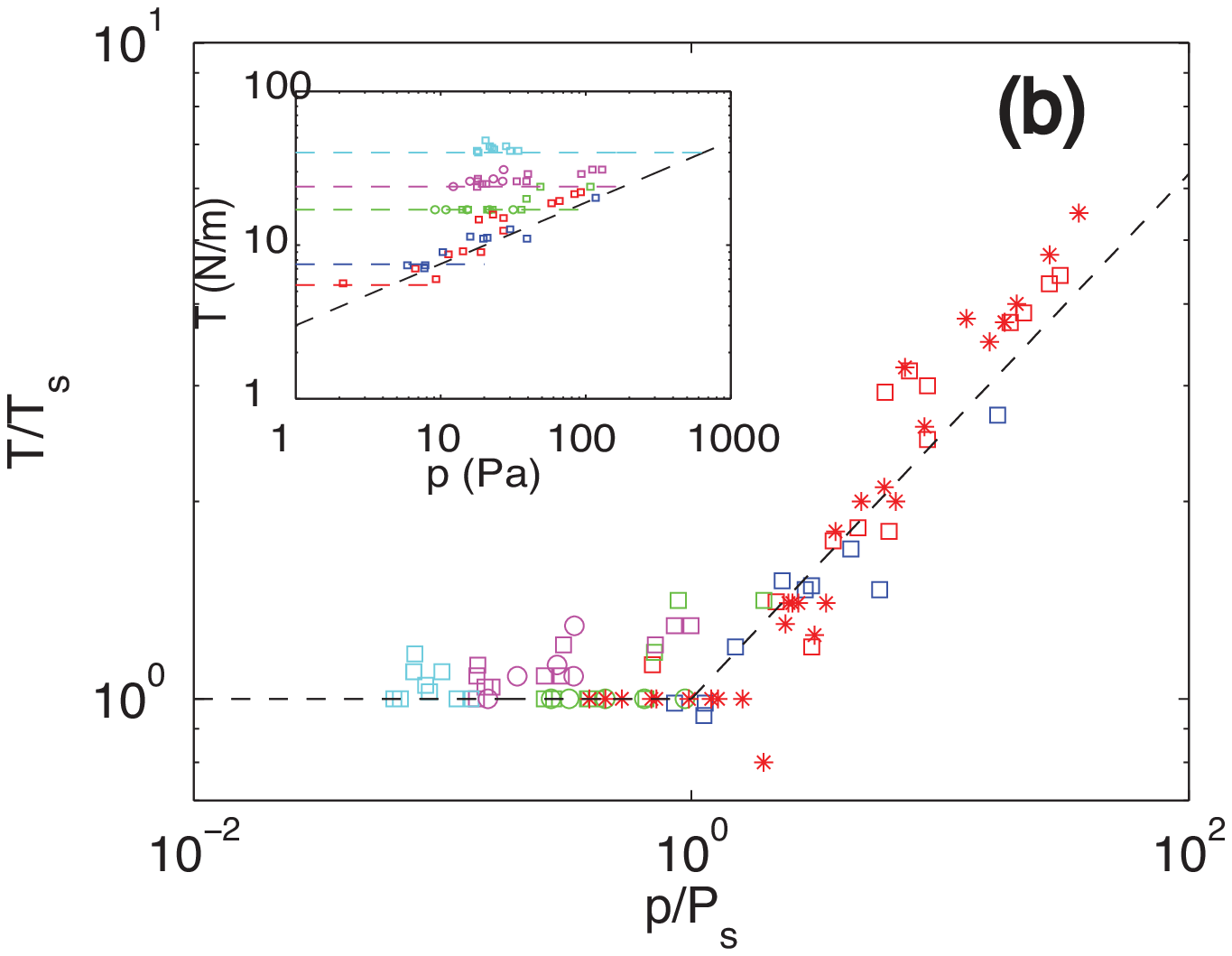}
\caption{(a) Main: Dispersion relations when the forcing amplitudes is increased (from top to bottom): $p$ goes from 3 (top) to 100 (bottom) Pa. $P_s=0$. Dashed lines correspond to $\omega^2=\frac{D}{\rho}k^5+\frac{T}{\rho}k^3$ with $T$ the only fitting parameter: $T=4$ to 40 N/m (from top to bottom). Inset: same figure with $P_s=130$ Pa ($T_s=22$ N/m), with $10 \leq p \leq 130$ Pa. Dashed line, best fit: $T=22$ N/m (b) Main: $T/T_s$ as a function of $p/P_s$, for various $P_s=$ ({\color{red}$\square$}) $P_s^0$, ({\color{blue} $\square$}) 7, ({\color{green}$\square$}, {\color{green}$\circ$}) 55, ({\color{magenta}$\square$}, {\color{magenta}$\circ$}) 130, ({\color{cyan}$\square$}, {\color{cyan}$\circ$}) 274 Pa for $R=10$ cm, and  $P_s=P_s^0=4$ ({\color{red}$\star$})  Pa for the $R=20$ cm. These values correspond, from Eq.\ (\ref{stth}), to $T_s=$4, 9, 17, 22, 39 N/m, and $T_s=5$ N/m, respectively. Dashed lines correspond to $T/T_s=1$ for $p/P_s<1$, and $T/T_s\sim (p/P_s)^{0.4}$ for $p/P_s>1$. Inset: same figure in dimensional units. For all colors, $(\square,\star)$ indicate depression and $(\circ)$ overpressure.}
 \label{fig03}
\end{center}
\end{figure}

Let us now introduce a dynamical pressure $p$ due to the wave field that will correspond to a mean ``dynamical'' tension $T$ on the interface. The dynamical pressure of the wave field on the interface is given by Eq. (\ref{eqpnl}) and is estimated as: $p(x,y,t)=\rho g \sigma_{\eta} + \rho\sigma_{v}^2/2$, where $\sigma_{\eta}$ and $\sigma_{v}$ are the rms value $\eta(x,y,t)$ and $v(x,y,t)$ such as $\sigma_X\equiv \frac{1}{\mathcal{T}}\int_0^{\mathcal{T}}\sqrt{\langle(X-\langle X\rangle)^2\rangle}dt$ with $\langle \cdot \rangle$ denotes 2D spatial averaging on ($x$,$y$), and $\mathcal{T}$ is the total recording time. Since the system is stationary, the time dependent term in Eq. \ref{eqpnl} vanishes. Moreover, we have checked that $p$ is proportional to the mean power $I$ injected by the wave-maker within the system. One finds $p=bIf_p$ for our range of forcing frequencies $f_p$, with $b$ a dimensional constant depending on the forcing shape (sinusoidal or random). Thus, $p$ will control the forcing amplitude in the following. One can now compare the influence of the hydrostatic  pressure, $P_s$, and of the dynamical one, $p$, on the dispersion relation of the waves. 

Figure \ref{fig03}a shows the dispersion relation extracted from the maxima of the spectrum $S_v(k,f)$ for each scales, for various forcing amplitudes, and for $P_s=0$. A  significant shift of the dispersion relation is observed when the forcing is increased. The theoretical dispersion relation $\omega^2=\frac{D}{\rho}k^5+\frac{T}{\rho}k^3$ well describes the data when the single fitting parameter $T$ varies from $T=4$ N/m (top curve) at the lowest forcing amplitude up to $T=40$ N/m (bottom curve) for the highest forcing one. This minimum value of $T$ is consistent with the static tension $T_s^0$ (with no water) only due to the sticking of the sheet on the vessel (see estimation in \S  \ref{setup}). The shift of the dispersion relation with the forcing is observed when $p>P_s^0$. For fixed $P_s$ and a forcing amplitudes, such that $p\lesssim P_s$, no shift is observed on the dispersion relation (see inset of Fig. \ref{fig03}a). We have then performed similar experiments for various applied $P_s$, and the value of a mean ``dynamical'' tension $T$ is deduced from the fit of the dispersion relation as in Fig. \ref{fig03}a. $T$ is plotted as a function of $p$, for various applied pressure $P_s$ (and thus various $T_s$, from eq. (\ref{stth})) in the inset of Fig. \ref{fig03}b, and in rescaled variables $T/T_s$ and $p/P_s$ in the main Fig. \ref{fig03}b. Fig. \ref{fig03}b shows the collapse of all data on a master curve displaying two regimes: for $p<P_s$, one has $T\simeq T_s$, meaning that the nonlinear effects of the waves are not strong enough to shift the dispersion relation (as in inset of Fig. \ref{fig03}a). On the other hand, when $p>P_s$, $T$ is found to increase with $p$ as $T\sim p^{0.4}$. The inset of Fig. \ref{fig03}b shows the unrescaled variables. When the forcing $p$ is increased, $T$ is first constant, equal to the static tension $T_s$ (horizontal dashed lines), and then follows the nonlinear law $T\sim p^{0.4}$ for various $P_s$. Note that when $P_s$ is applied, there is no effect of the sheet curvature on the wave propagation since the wavelengths are small compared to the radius of curvature.

\subsection{Sheet tension induced by nonlinear waves}

\begin{figure}
\begin{center}
\includegraphics[scale=0.6]{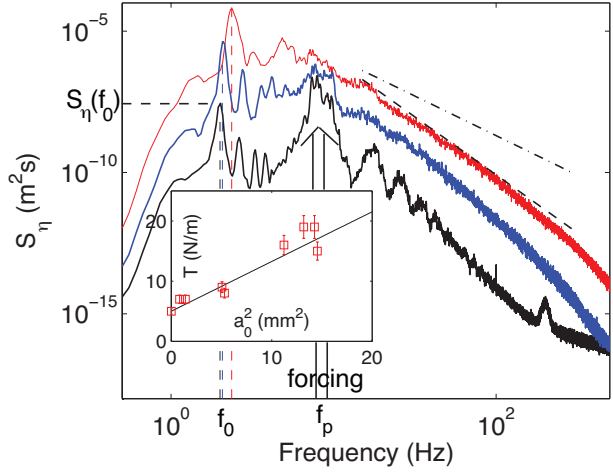}
\caption{Main: Frequency spectrum $S_{\eta}(f)$ of the wave height  for increasing forcing amplitudes (bottom to top). Single point temporal measurement. Forcing bandwidth: 7 - 10 Hz. $P_s=0$ and $T_s^0=4$ N/m. First vessel eigenmode ($f_0$, $k_0$), $f_0$ increases with the forcing (see vertical dashed lines). Dashed line has a slope $-4.1$. Dash-dot line has a slope $-17/6$. Inset: $T$ extracted from 2D measurement as a function of the amplitude $a_0$ of the vessel eigenmode at $f_0$. Solid line is the nonlinear model of Eq.\ (\ref{Tnl}) with $c=3$ the only fitting parameter. $10 \leq p \leq 40$ Pa.}
\label{fig04} 
\end{center}
\end{figure}

We now explain that the observed shift of the dispersion relation is due to the mean tension induced by the waves. Let us consider a 1D elastic string, fixed at both ends, subjected to an initial static force $F_s$ with a Young modulus $E'$ and a cross section $\Sigma$. When a relatively large transverse displacement of amplitude $a_0$ is imposed, at wave number $k_0$, the string elongates and longitudinal waves are generated due to a geometrical nonlinearity. The latter modulate the tension force of the string as  $F=F_s + F_{nl}$, with $F_{nl}\sim E'\Sigma(a_0 k_0)^2$ (\cite{Morse,Fletcher}). $F_{nl}$ is a nonlinear tension due to the coupling between the longitudinal and the transverse motions of the string. The nonlinear force $\sim (a_0 k_0)^2$ is similar to the one involved in the radiation pressure in acoustics (\cite{Morse}). Let us now apply abruptly this result for a 1D elastic string to a 2D elastic sheet. The tension stress $T$ of the sheet thus reads 
\begin{equation}
T=T_s+cEh(a_0k_0)^2 ,
\label{Tnl}
\end{equation}
with $h$ the sheet thickness, $E$ the Young modulus, and $c$ a geometry-dependent dimensionless constant. Eq.\ (\ref{Tnl}) is consistent with our above observations of increasing $T$ with the forcing amplitude, if one single wave mode is experimentally involved. To identify its wave number $k_0$ and determine its amplitude $a_0$, long time recording of the wave height $\eta(t)$ at a single location of the sheet is performed to resolve the low frequency response of the system. Figure \ref{fig04}a shows the spectrum of $\eta(t)$ for different forcing amplitudes, and for $P_s=0$. At low forcing (bottom curve), the forcing response (7 - 10 Hz) is observed as well as their harmonics. A peak is also visible at low frequency $f_0$ near 2 Hz (see dashed black line). When the forcing is increased, the high frequency part of the spectrum displays a frequency power-law, whereas the low-frequency part shows that the peak initially at $f_0$ grows strongly in amplitude, and is slightly increased in frequency (see vertical dashed lines). This peak amplitude is found to grow nonlinearly with the forcing, and becomes the most energetic frequency at high enough forcing (see both top curves). The peak frequency $f_0$ corresponds to the first antisymmetric eigenmode, $k_{0}$, of a circular sheet determined by the zero of Bessel function of the first kind (\cite{Morse}). Indeed, one finds $k_{0}=19$ m $^{-1}$ (i.e., $\lambda_0=33$ cm) for $R=20$ cm, and thus $f_{0}\approx2$ Hz using Eq. (\ref{LDR}) with $T=T_s^0$.

The frequency shift of the sheet eigenmode with the amplitude is consistent with the increase of $T$ observed above. Indeed, the values of $T$ (for the same forcing parameters as in the one point measurements) are deduced from 2D space-time measurements (see above). The amplitude $a_0$ of the mode $k_0$ is extracted from Fig. \ref{fig04}a, using the relation $a_0^2(f_0)=\int_{f_0}^{f_0+\delta f}S_{\eta}(f)df$, with $\delta f=0.06$ Hz the spectrum frequency resolution. When plotting the tension $T$ as a function of $a_0$ as in the inset of Fig. \ref{fig04}a, one finds that $T = T_s^0 + c' a_0^2$, with $c'=c\ Eh$ and $T_s^0=4$ N/m, in agreement with Eq.\ (\ref{Tnl}). As the eigenvalue frequency is much less that the elastic wave one ($5\lesssim f \lesssim 500$ Hz), the eigenmode oscillations is quasi-static with respect to the wave propagation. To sum up, the nonlinear quasi-static oscillations of the fundamental eigenmode of the sheet induce a mean additional tension on the whole sheet.

\subsection{Scaling of tension with pressure: Dynamical case}

In \S \ref{shift}, we have experimentally shown that $T\sim p^{0.4}$. This scaling is explained as follows. First, assume a tension wave (of wave number $k$ and amplitude $a$) propagating on a 1D string. The power $I$ due this transverse wave is $I\sim F\cdot v$, with $F$ the tension force and $v$ the velocity of the wave. Assuming $v=\partial a/\partial t \sim a\omega$, and $F=T\partial a/\partial x \sim Tak$, one has thus $I=Ta^2\omega k$. For a large enough $a$, only a dynamical tension $T\gg T_s$ acts on the string. Using Eq.\  (\ref{Tnl}), that is  $T\sim (ak)^2$, one has $I=T^2\omega /k$.  For the dispersion relation of a 1D string $\omega=\sqrt{T/\rho}k$, one finds $T\sim I^{2/5}\rho^{1/5}$. Similarly, for the dispersion relation of pure tension waves on a 2D floating sheet, $\omega^2=(T/\rho) k^3$, and one has $T \sim I^{2/5}(\rho/k)^{1/5}$. Finally, using the experimental observation $I\sim p$, one gets $T \sim p^{2/5}(\rho/k)^{1/5}$ in good agreement with our observations in $T \sim p^{0.4}$ in Fig.\ \ref{fig03} and \S \ref{shift}.

\section{Wave turbulence\label{wt}}
Here, we study the regime of wave turbulence when the system is subjected to a random noise forcing at large scale.

\begin{figure}
\begin{center}
\includegraphics[scale=0.6]{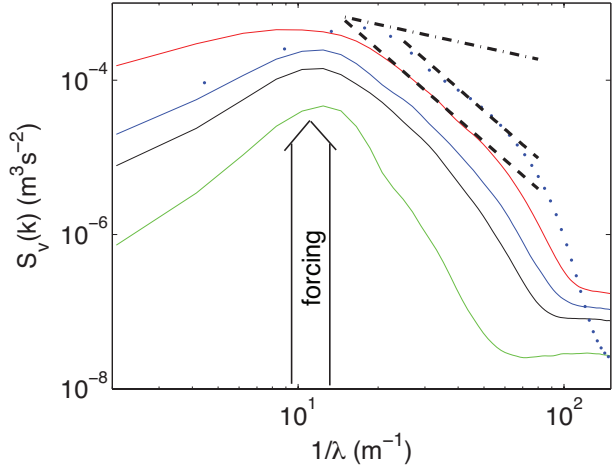}
\caption{Spatial spectrum $S_v(k)$ of the wave transverse velocity for increasing forcing amplitudes (bottom to top). Solid lines: large vessel, forcing bandwidth: 7 - 10 Hz. Dotted line: small vessel, forcing bandwidth: 15 - 20 Hz. Dashed lines have a slope $-3$. Dash-dot line has a slope $-3/4$. $10 \leq p \leq 40$ Pa.}
\label{figwt} 
\end{center}
\end{figure}

\subsection{Wave power spectrum}
Figure \ref{fig04} shows the power spectrum of the wave height  $S_{\eta}(f)$ as a function of frequency for different forcing amplitudes. At low forcing, the spectrum exhibits peaks at the forcing frequencies and at their harmonics. When the forcing is high enough, $S_{\eta}(f)$ displays a power law $\sim f^{-4.1}$ on one order of magnitude in frequency (see dashed line), and is thus found to be scale-invariant as expected for a wave turbulence regime. This power-law exponent is far from the $-17/6\simeq -2.8$ exponent (see solid line) predicted by dimensional analysis (see Eq. (\ref{DAtime})). 

The spatial spectrum $S_v(k=||\mathbf{k}||)$ of the transverse velocity is computed by integrated the space-time spectrum of $S_v(\mathbf{k},f)$ over all the directions of $\mathbf{k}$ and over $f$. Figure \ref{figwt} shows $S_v(k)$ when the forcing is increased and for two sheet sizes. At low forcing amplitude, the forcing response is mainly observed generating a very steep spectrum. When the forcing is further increased, energy is redistributed to higher frequencies, leading to a less steep spectrum and a regime roughly in power law of the spatial scales on less than one decade ($1/\lambda \approx 20$ to 80 $m^{-1}$). This power-law spectrum scales as $S_v(k)\sim k^{-\alpha}$, with $\alpha=3\pm0.2$. $\alpha$ is found to be independent of the sheet size and the forcing bandwidth used, but is far from $-3/4$ (see dotted-dashed line), the one predicted by dimensional analysis (see Eq. \ref{DAtension}). Note that a power law is only reached when $p/P_S \gg 1$, i.e. when nonlinear effects become important. This observation can be done whatever the static pressure. The presence of bending waves does not explain for this discrepancy since bending wave turbulence is predicted to scale as $k^{-1/4}$ (see Eq. \ref{DAflex}). Finally both experimental spectrum scalings in space and in frequency suggest that the change of variable $k \longleftrightarrow f$ using the linear dispersion relation to estimate $S_{v}(f)$ from $S_{v}(k)$ is consistent. Indeed, using $S_{v}(\omega)=S_{v}(k)/(d\omega/dk)$ and $S_{v}(k) \sim k^{-3}$ found experimentally, one finds $S_{v}(f)\sim f^{-7/3}$, thus $S_{\eta} \sim f^{-13/3}$ for pure tension waves ($\omega(k)\sim k^{3/2}$). This estimated exponent is coherent with the one find experimentally, $S_{\eta} \sim f^{-4.1}$. 

\subsection{Forcing induced anisotropy\label{isotropy}}

\begin{figure}
\begin{center}
\includegraphics[scale=0.3]{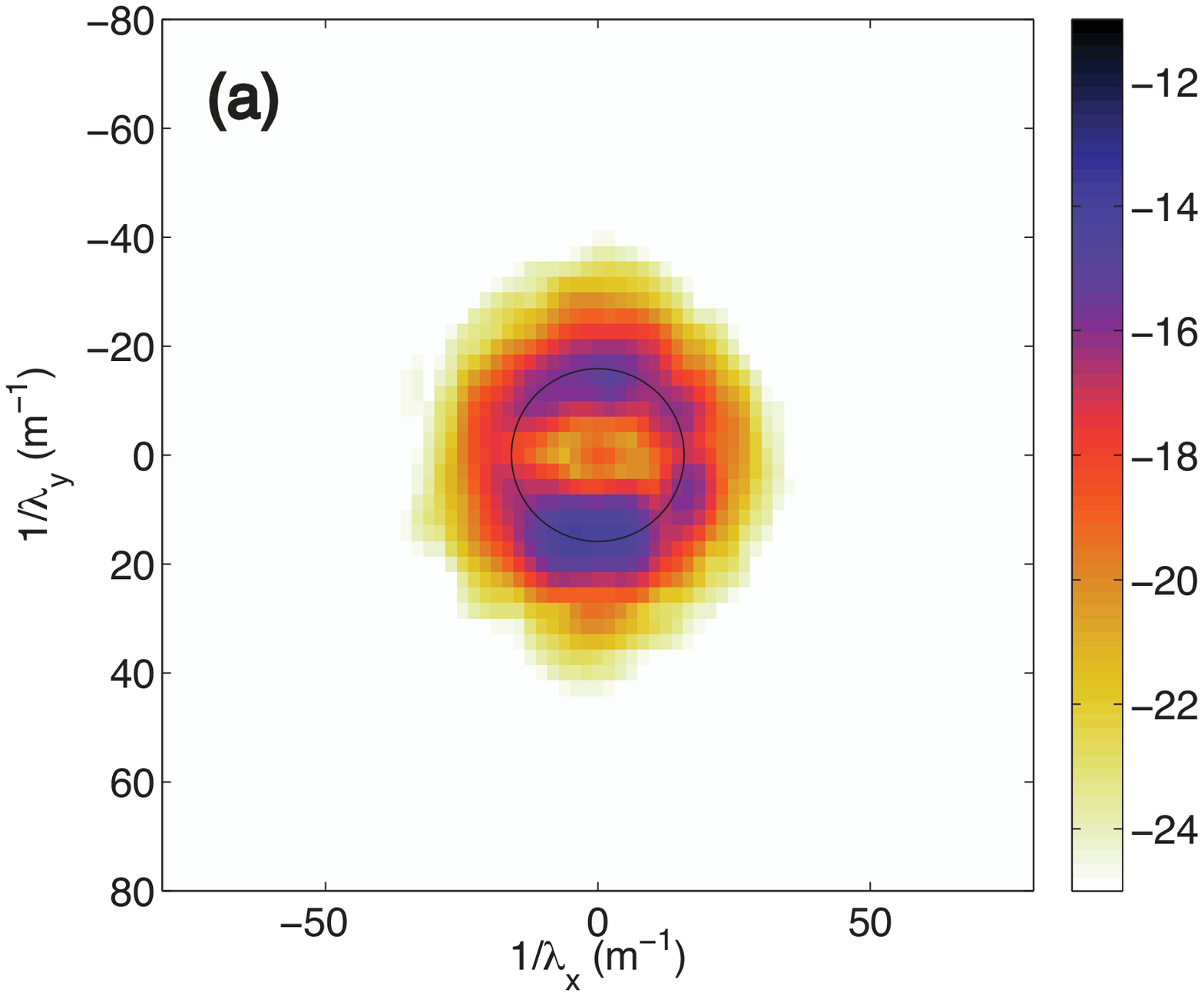}
\includegraphics[scale=0.3]{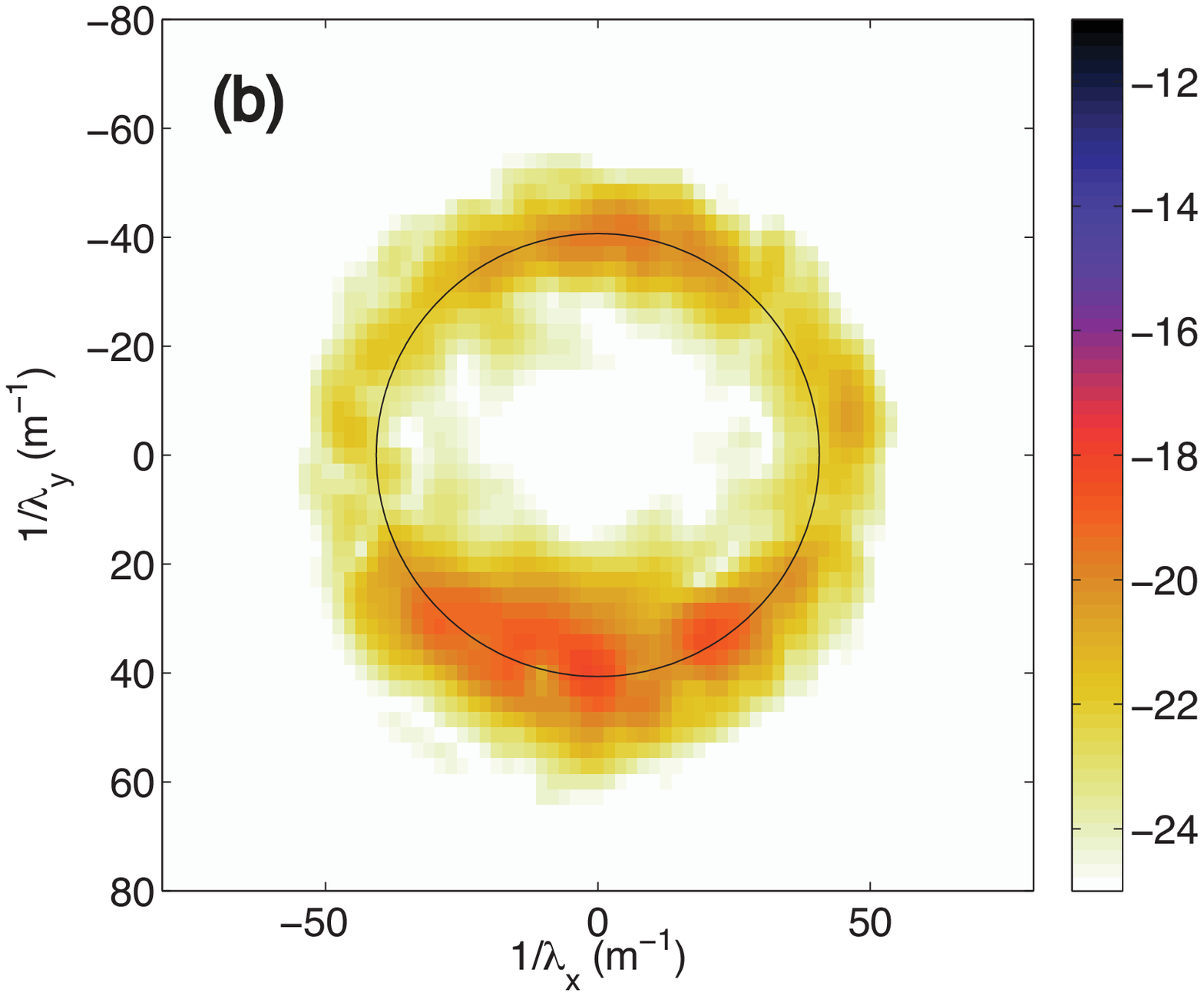}\\
\includegraphics[scale=0.3]{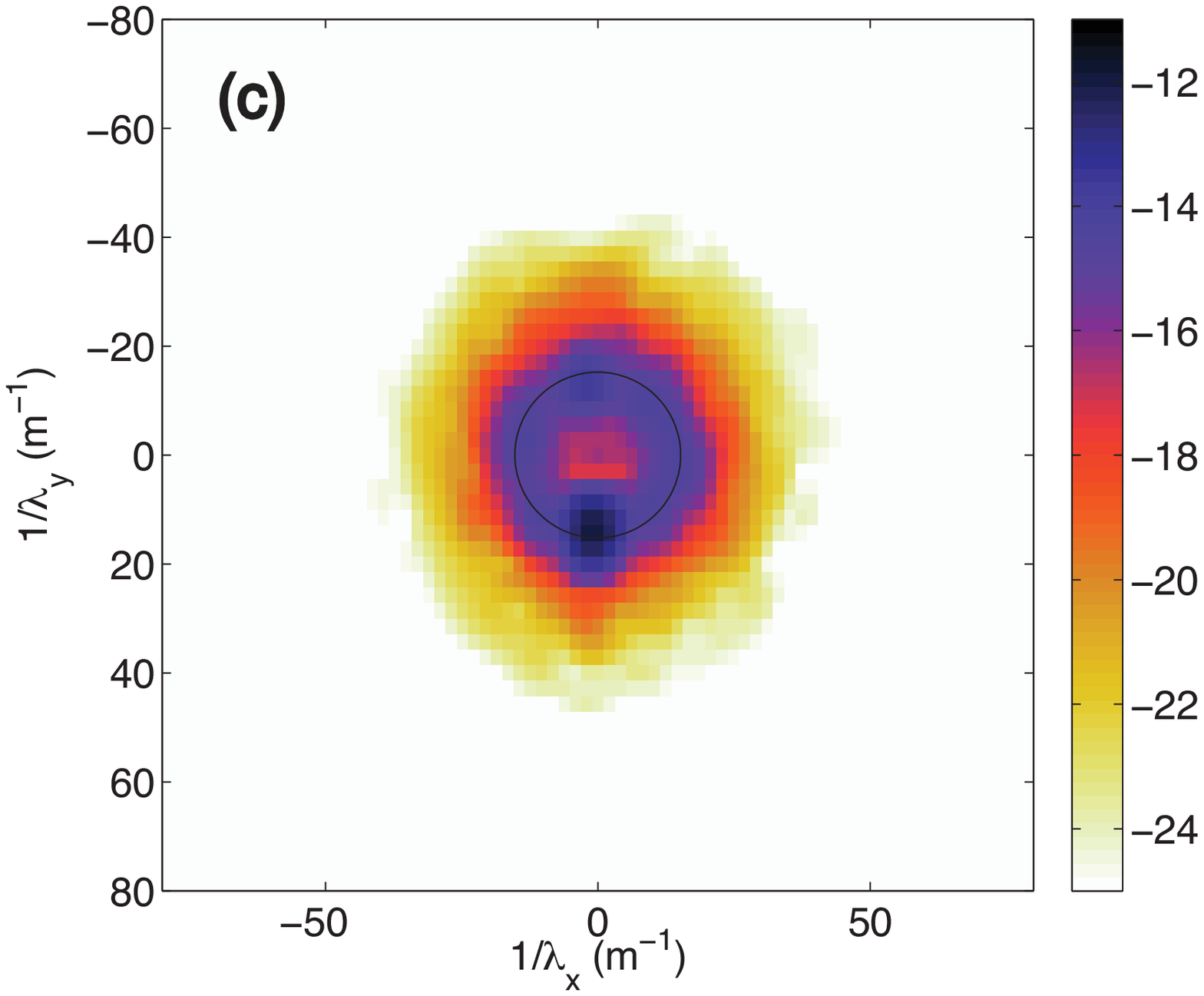}
\includegraphics[scale=0.3]{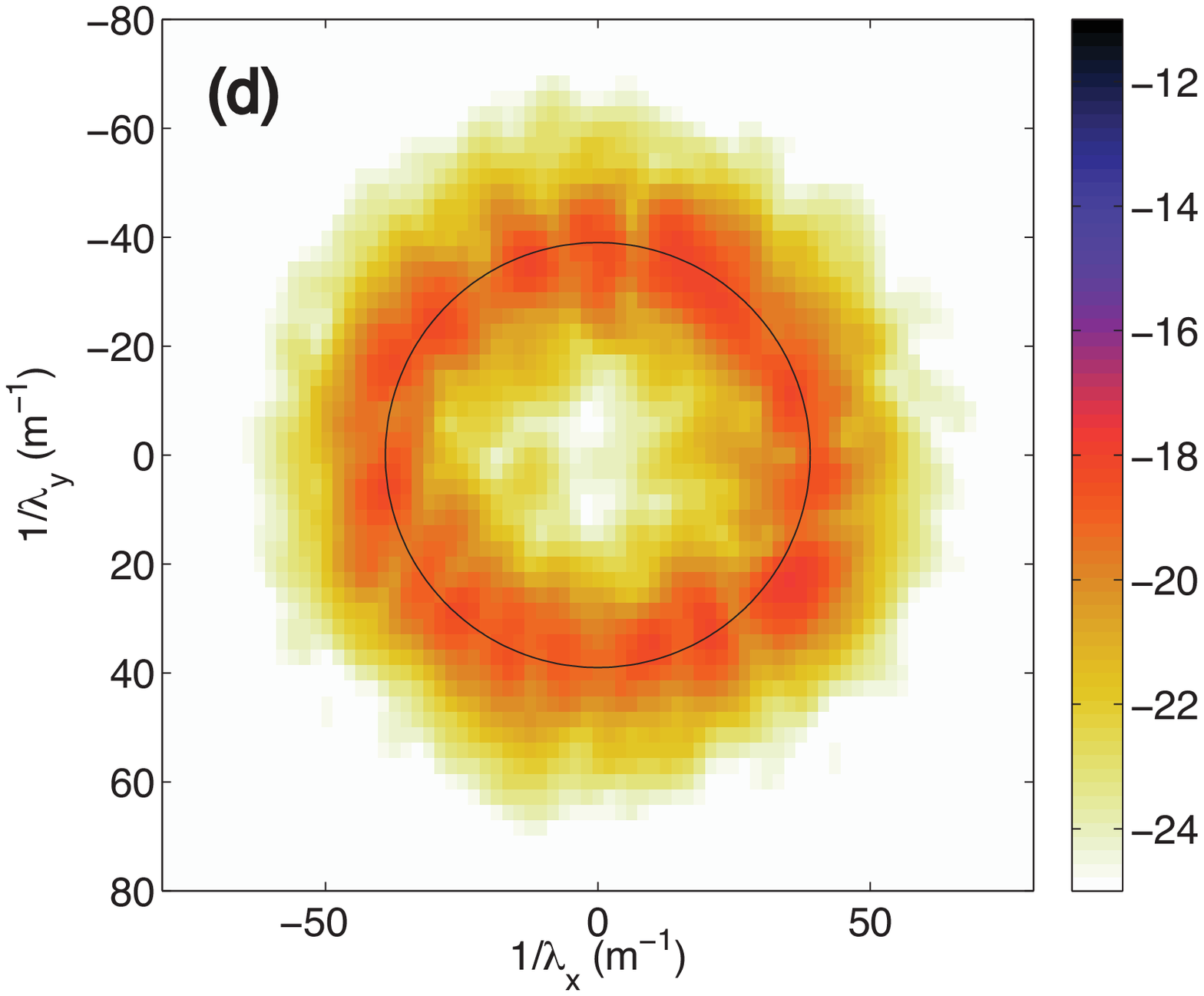}
\end{center}
\caption{Space spectrum $S_v(k_x,k_y)$ at a fixed frequency $f^*=$ 20 (a, c) or 80 (b, d) Hz, for a moderate (a, b) and strong forcing (c, d). Energy is localized on a circle of radius $k^*$ given by the dispersion relation. At low frequency (a, c), the maximum of energy is localized is the forcing direction ($y$-axis). At high forcing and high frequency (d) the spectrum is more isotropic.}
\label{fig05}
\end{figure}

Isotropy of the wave field is assumed by wave turbulence theory (\cite{NewellReview}). The spatial spectrum $S_v(k_x,k_y,f^*)$ at a given frequency $f^*$ is shown in Fig. \ref{fig05} for moderate and strong forcing. The wave energy is localized around a ring in the Fourier space, which radius $k^*=k(f^*)$ is given by the linear dispersion relation of Eq. (\ref{LDR}). At low frequency ($f^*=20$ Hz),  energy is spread in all directions (see Figs. \ref{fig05} a, c), the maximum of energy being near the forcing direction ($y$-axis). Thus, the forcing induces anisotropy at low frequency. At higher frequency ($f^*=80$ Hz), this anisotropy is still observed for moderate forcing (see Fig. \ref{fig05}b), and becomes much more isotropic at strong forcing (see Fig. \ref{fig05}d). This is likely due to nonlinear interactions and multiple reflexions on the vessel boundary that enables energy transfers at small scale and redistribution in all directions. Different frequencies or forcing conditions lead to the same conclusions: the forcing induces anisotropy at large scale, and isotropy is reached at small scale for strong enough forcing.

\begin{figure}
\begin{center}
\includegraphics[scale=0.35]{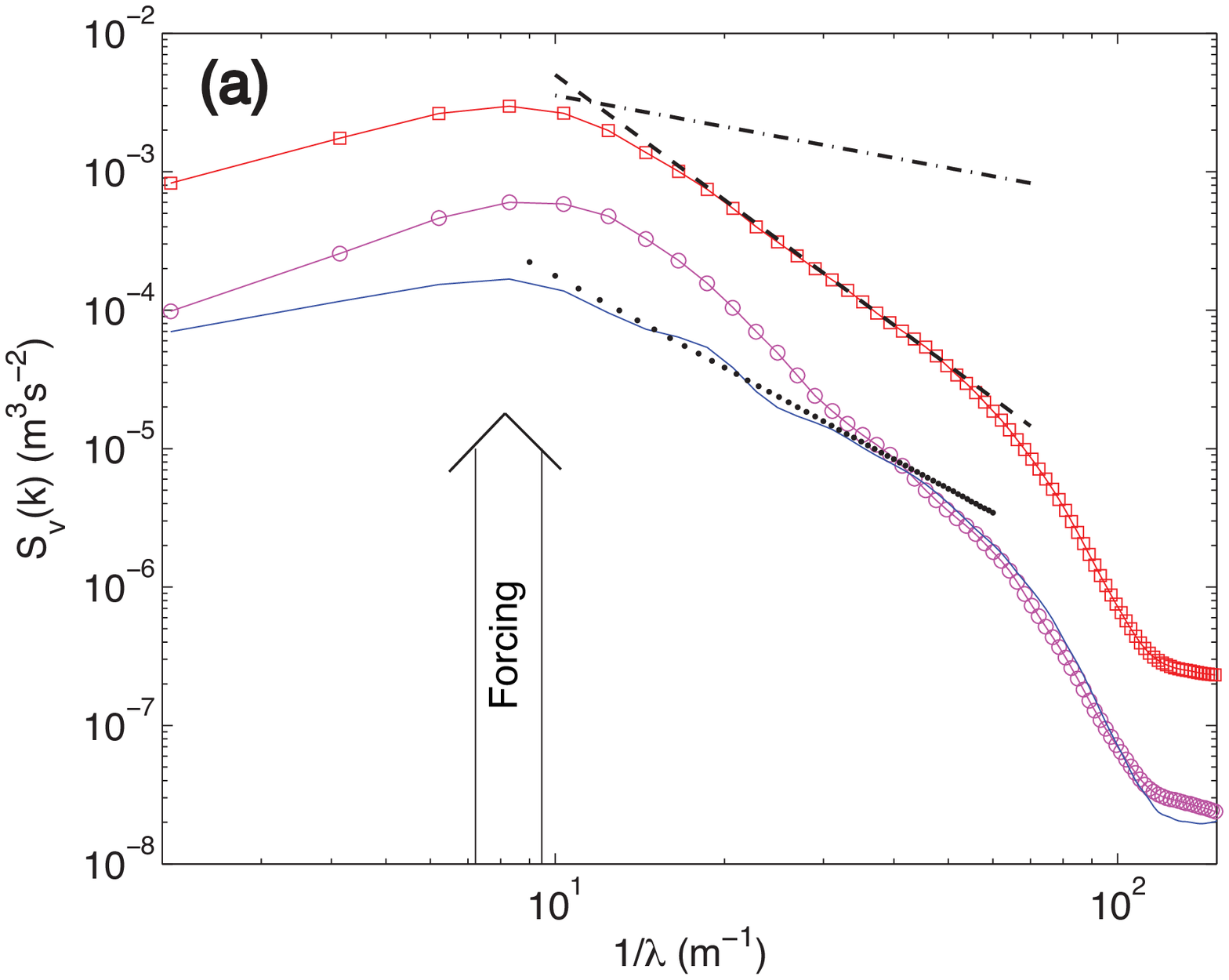}\includegraphics[scale=0.35]{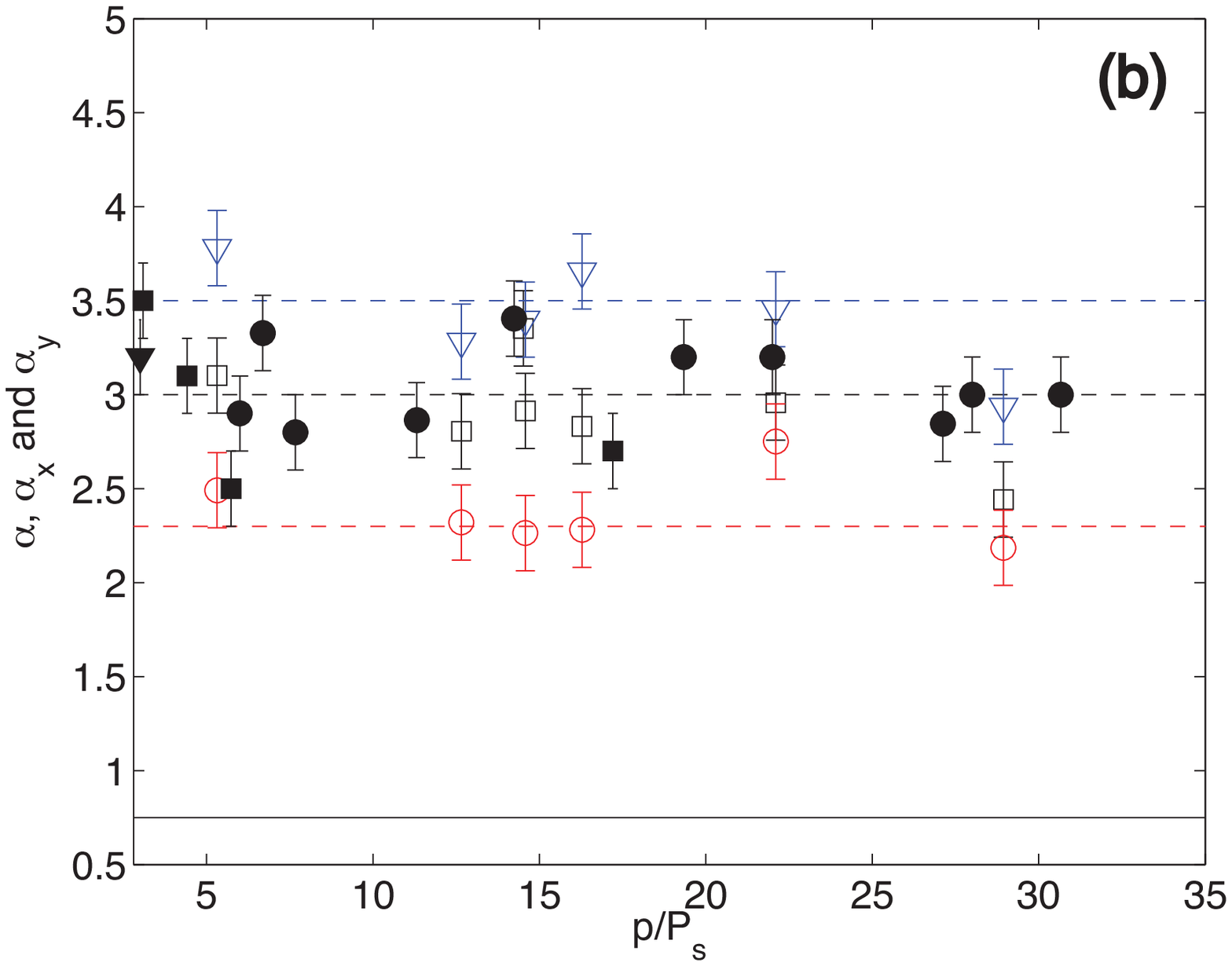}
\caption{(a) Spatial spectrum $S_v(k)$ of the wave transverse velocity. ($\square -$) $S_v(k)$, ($\circ -$) $S_v^{y}(k)$, and ($-$) $S_v^{x}(k)$. Dashed line has a slope $-3$. Dash-dot line has a slope $-3/4$. Dotted line has a slope $-2.3$. Large vessel, forcing bandwidth: 7 - 10 Hz. $p=70$ Pa. (b) Power-law exponents. Large vessel: ($\square$) $\alpha$, ($\circ$) $\alpha_x$ and ($\triangledown$) $\alpha_y$, $P_s=P_s^0=5$ Pa. Small vessel: $\alpha$: ($\bullet$) $P_s=P_s^0=3$ Pa, ($\blacksquare$) $P_s=7$ Pa, and ($\blacktriangledown$) $P_s=55$ Pa.}
\label{fig06} 
\end{center}
\end{figure}

The spatial spectrum $S_v(k)$ of Fig. \ref{figwt} is obtained by integrating $S_v(\mathbf k)$ over all directions of $\mathbf k$. It thus averages the spectrum components in the forcing direction ($y$-axis) and in other directions. By integrating $S_v(\mathbf k)$ over the $y$-axis, one gets its component in forcing direction $S_v^y(k)$. Similarly, by integrating $S_v(\mathbf k)$ over the $x$-axis, one has its component in the normal direction to the forcing $S_v^x(k)$. The angle partial integration is performed with a 0.5 rad tolerance around the direction ($x$ or $y$). Figure \ref{fig06}(a) shows $S_v(k)$, $S_v^x(k)$, and $S^y_v(k)$. As the energy maximum is localized in the $y$ direction (see above), the maximum of $S_v^y(k)$ is almost one order of magnitude greater than $S_v^x(k)$. Moreover, all spectra display power laws, say $S_v^x(k)\sim k^{-\alpha_x}$ and $S_v^y(k)\sim k^{-\alpha_y}$, with $\alpha_x \neq \alpha_y$. More precisely, one finds $\alpha=3\pm0.3$, $\alpha_x=2.3\pm0.3$ and $\alpha_y=3.5\pm0.5$ each being roughly independent of the forcing, $p/P_s$, as shown in Fig. \ref{fig06}b, but all far from the $-3/4$ prediction of Eq. (\ref{DAtension}). 

When now computing $S_v(k,f)$ either on one quarter or on the whole spatial window leads to similar results: the wave energy is redistributed on a curve described by the theoretical dispersion relation with the same value of $T$. Computing $S_v(\mathbf k,f)$ integrated over all directions of $\mathbf k$ or only for some chosen directions, gives also the same value of $T$. Although the wave field is not isotropic at large scale, the sheet tension is thus isotropic and homogeneous as assumed theoretically in $\S$ \ref{TvsP}.

To resume, we observe an anisotropy of the wave field at large scale induced by the forcing. The spectrum computed either in the forcing direction or in the normal one leads to different power law scalings with $k$, the one in the forcing direction being the steepest, and both are in disagreement with the prediction of Eq. (\ref{DAtension}). When performing experiments with two wave makers with normal directions to each other:  $S_v(k)$, $S_v^x(k)$, and $S^y_v(k)$ are found to be similar with the same power law scaling in $k^{-3}$, still far from the $k^{-3/4}$ prediction. Thus, the anisotropy is not the main origin for this discrepancy between the theoretical and experimental exponents.

\subsection{Role of the sheet finite size}
When the vessel diameter is reduced by a factor 2, a power-law spectrum, $S_v(k)\sim k^{-\alpha}$ is still observed as shown in Fig. \ref{figwt}. Fig. \ref{fig06}b shows $\alpha$ as a function of $p/P_s$ for the small and large vessels with $\alpha=3\pm0.3$ whatever the vessel size used here. This value of $\alpha$ is also independent on the sheet tension for $3 \leq P_s \leq 60$ Pa for which power law spectrum is observed. The regime of wave turbulence is thus independent of the vessel size and of the static sheet tension whatever the forcing in the range $5 \leq p/P_s \leq 30$. 

\subsection{Time scale separation}
Let us now consider the typical time scales involved in our experiment. Wave turbulence theory assumes a time scale separation $\tau_l(k) \ll \tau_{nl}(k) \ll \tau_d(k)$, between the linear propagation time, $\tau_l$, the nonlinear interaction time, $\tau_{nl}$, and the dissipation time, $\tau_d$. The linear propagation time is $\tau_l=1/\omega(k)$, with $\omega(k)$ the linear dispersion relation of Eq. (\ref{LDR}). $\tau_d(k)$ is determined by performing freely decaying experiments (\cite{Miquel2011}). $\tau_{nl}(k)$ is related to the broadening of the dispersion relation and can be  experimentally estimated by the width of the spatio-temporal spectrum $S_v(k,f)$ (\cite{NazarenkoBook}, \cite{Miquel2012}). 

\subsubsection{Dissipation time\label{td}}
The dissipation time, $\tau_d(k)$, at each scale $k$ is determined experimentally by performing freely decaying experiments. The protocol is similar to the one of recent works on gravity-capillary wave turbulence (\cite{Deike2012}) or on flexural wave turbulence on a plate (\cite{Miquel2011}). First, surface waves are generated to reach a stationary wave turbulence state (see $\S$ \ref{setup}). Then the forcing is stopped at $t=0$, and the temporal decay of the spatial wave field is recorded during a time $\mathcal{T}=3$ s. The experiment is automatically iterated 10 times to improve statistics, and results are averaged (ensemble average denoted by $\langle \cdot \rangle$). To analyze the different steps of the decay, the wave velocity spectrum is computed over short-time intervals $[t,t+\delta t]$ with $\delta t= 0.1$ s. Experiments are performed in the large vessel, and similar results is obtained whatever the forcing. 

\begin{figure}
\centering
\includegraphics[scale=0.35]{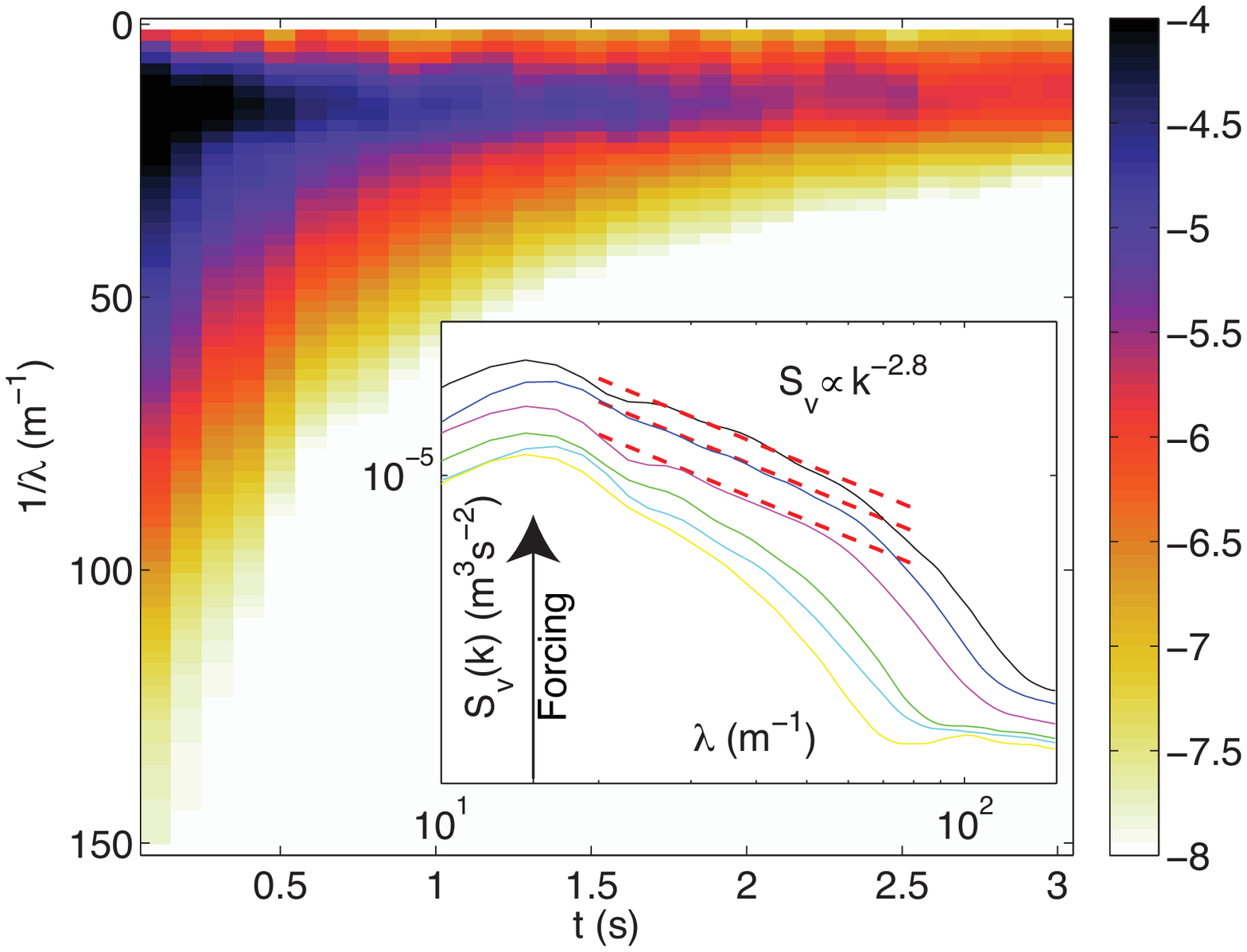}\includegraphics[scale=0.35]{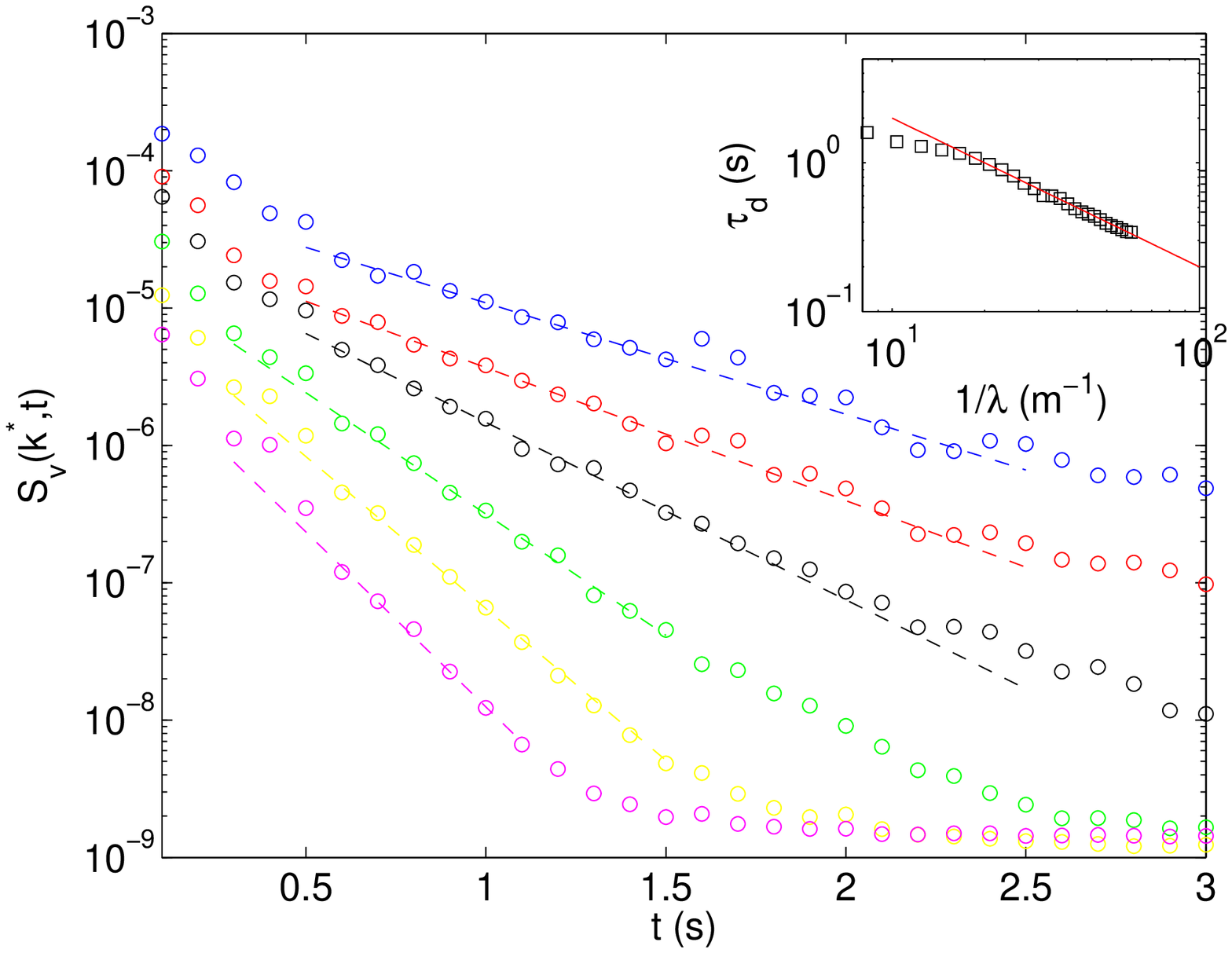}
\caption{Temporal decay of the space-time spectrum for a sinusoidal forcing at $f_p=20$ Hz. Left: main: $\langle \log{S_v(k,t)} \rangle$. Inset: space spectrum $\langle S_v(k,t^*) \rangle$ at different times $t^*=0.1$, 0.2, 0.4, 0.6, 0.8 and 1 s (from top to bottom). Dashed lines have a slope $-2.8$. Right: main: ($\circ$) Temporal decay of the spectrum Fourier component $\langle S_v(k^*,t) \rangle$ at mode $k^*=116.9$, 143.2, 182.2, 246.3, 312.3, and 377.6 m$^{-1}$ (from top to bottom). Dashed lines are $\exp{[-t/\tau_d(k^*)]}$ with $\tau_d(k^*)$ fitted from the data (from top to bottom). Inset: $\tau_d$ versus $k$. ($\square$) experiments, and ($-$) best fit $\tau_d=Ak^{-1}$ with $A=3.2$ s/m$^{-1}$.}
\label{dec1}
\end{figure}

The temporal decay of the space-time power spectrum $\langle S_v(k,t) \rangle$ is displayed in Fig. \ref{dec1}(left) and shows that small scales are first dissipated. The inset shows the temporal evolution of the spatial spectrum $\langle S_{v}(k,t^{*}) \rangle$ at different decay times $t^{*}$. It corresponds to different vertical sections of the space-time spectrum of the main Fig. \ref{dec1}(left). At the beginning of the decay ($t \lesssim 0.5$ s), the spectrum shape is conserved (see top curves in the inset Fig. \ref{dec1}(left)). For $t \lesssim 0.5$ s, one finds roughly $\langle S_{v}(1/\lambda,t^{*}) \rangle \sim k^{-2.8}$ within the inertial range. This exponent is thus close to the one found in the stationary state (see $\S$ \ref{isotropy}). This self-similar temporal decay of the wave spectrum has been predicted theoretically (\cite{ZakharovBook}), and has been observed for decaying capillary wave turbulence (\cite{Deike2012}), and decaying flexural wave turbulence on a plate (\cite{Miquel2011}). For $t \gtrsim 0.5$ s, no power law regime is observed, meaning that the wave turbulence regime stops in favor of a purely dissipative regime.

Figure \ref{dec1} (right) shows the temporal evolution of the amplitude of the Fourier modes $\langle S_{v}(k^*,t)\rangle$ at different $k^*$. It corresponds to different horizontal sections of the space-time spectrum of the main Fig. \ref{dec1}(left). For $t \gtrsim 0.5$ s, each Fourier mode is found to decay exponentially in time, $\langle S_v(k^*,t) \rangle \sim \exp{[-2t/\tau_d(k^*)]}$ with a typical time $\tau_d(k^*)$ given by the best fit of the experimental data. $\tau_d(k^*)$ is displayed in inset of Fig. \ref{dec1} (right). One finds $\tau_d(k^*) \sim 1/k^*$ meaning that small scales dissipates more rapidly than large ones. By using temporal measurement in one spatial point, one gets the frequency scaling, $\tau_d(f) \sim f^{-0.7}$ consistently with $\tau_d(k) \sim k^{-1}$ since $\omega^2=\frac{T}{\rho}k^3$. Similar scaling laws are observed whatever the initial static tension used. Even if the physical origin of the dissipation is not known, an important result of this part is that dissipation occurs at all scales within the inertial range in contrary to the hypothesis of wave turbulence theory. 

\subsubsection{Nonlinear interaction time\label{tnl}} 
\begin{figure}
\centering
\includegraphics[scale=0.35]{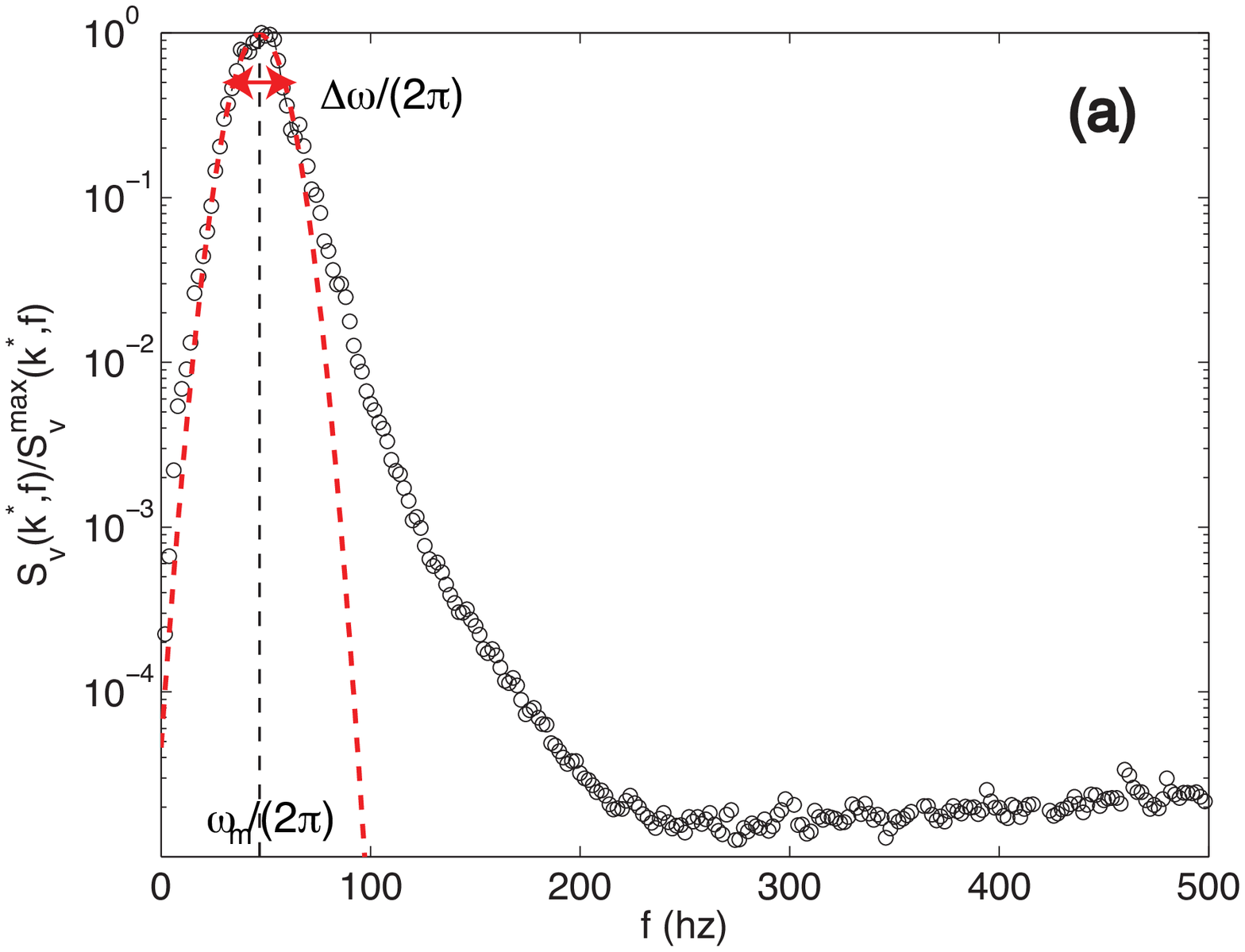}\includegraphics[scale=0.35]{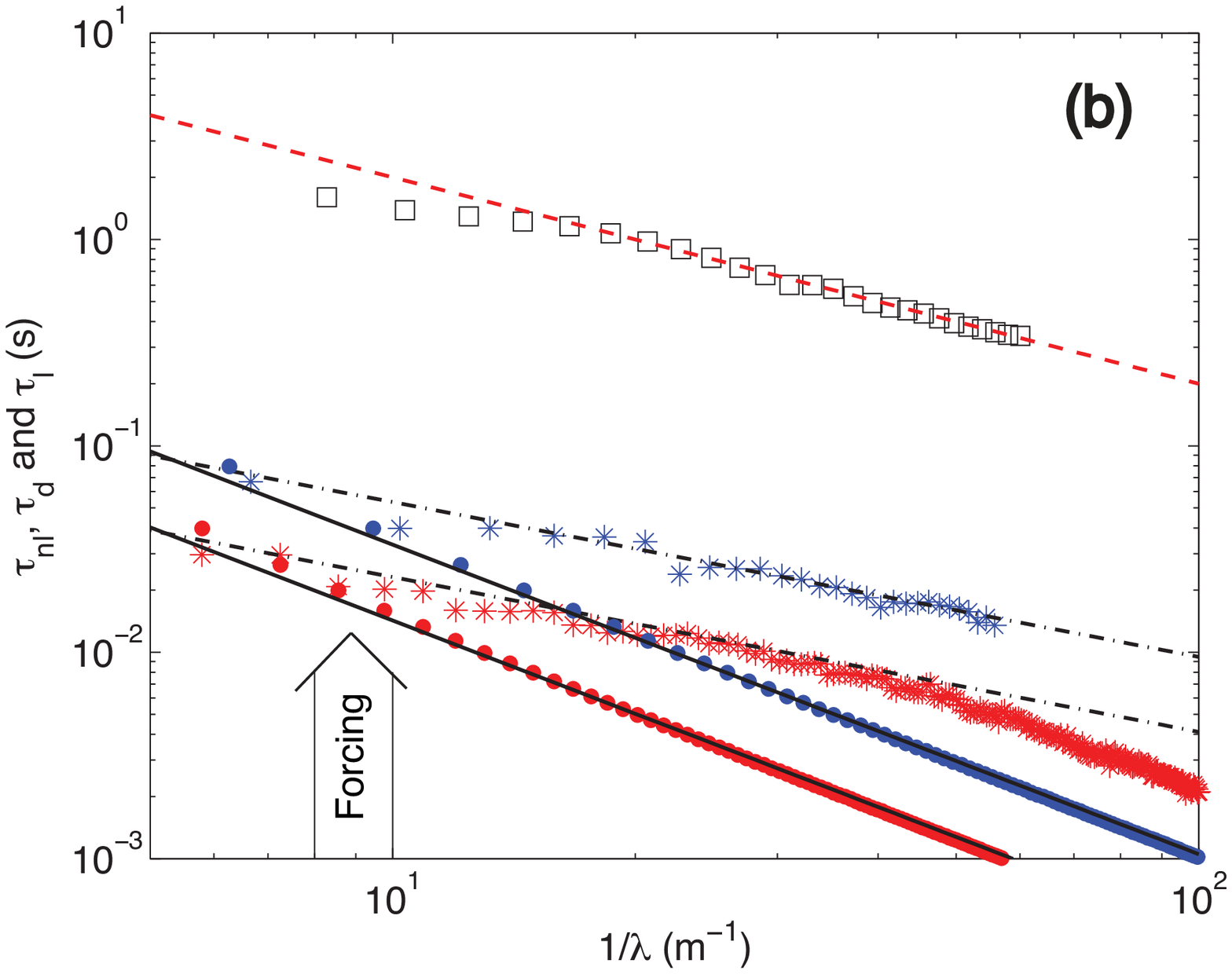}
\caption{(a) ($\circ$) Spectrum $S_v(k^*,f)$ normalized by its maximum for $k^*=2\pi \times 29$ m$^{-1}$. ($--$) Gaussian fit:  $\exp{[-(\omega -\omega^*)^2/\Delta\omega^2]}$ with $\omega^*=2\pi \times 47$ s$^{-1}$, and $\Delta\omega=2\pi  \times14.8$ s$^{-1}$. (b) Typical time scales:  ($*$) $\tau_{nl}$ and ($\bullet$) $\tau_l$ both for low (top (blue)) and strong (bottom (red)) forcings, and ($\square$) $\tau_d$. Dashed line: best fit $\tau_d \sim 1/k$. Dash-dot lines: $\tau_{nl} \sim k^{-3/4}$ from Eq. (\ref{tnl_dim}). Solid lines: $\tau_{l} \sim k^{-3/2}$ (tensional waves). Low forcing is $p=2$ Pa, and $T=4$ N/m. High forcing is $p=145$ Pa, and $T=17$ N/m.}
\label{tnl1}
\end{figure}

As shown previously in Figure \ref{fig02}, the spatio-temporal spectrum $S_v(k,f)$ broadens around the linear dispersion relation curve when the forcing is increased. The nonlinear time scale $\tau_{nl}(k)$ is related to this broadening (\cite{NazarenkoBook}, \cite{Miquel2012}). It is estimated by $\tau_{nl}(k)=1/\Delta \omega(k)$, the inverse of the width $\Delta \omega(k)$ of spectrum $S_v(k,f)$ at a given wave number $k$. As shown in Fig. \ref{tnl1}(a), $\Delta \omega(k)$ is extracted from $S_v(k,f)$ by the rms value of a Gaussian fit with respect to $k$ at a fixed $\omega^*$. Then, iterating this protocol to all $k$ allow us to determine $\tau_{nl}(k)$.  Fig. \ref{tnl1}(b) shows that $\tau_{nl}(k)$ is found to scale as $\tau_{nl}\sim k^{-3/4}$ for low and high forcings, in well agreement with the prediction of Eq. (\ref{tnl_dim}). When the forcing is increased, the degree of nonlinearity increases, and $\tau_{nl}$ is found to decrease. This is consistent with the theoretical scaling $\tau_{nl}\sim \epsilon^{-1/2}$ of Eq. (\ref{tnl_dim}).

\subsubsection{Time scale separation} 
 We can now take stock of our results to compare the evolution of the linear time $\tau_l(k)$, the nonlinear time $\tau_{nl}(k)$, and the dissipation time $\tau_d(k)$ across the scales. One has experimentally found $\tau_{nl}\sim k^{-3/4}$ (see $\S$ \ref{tnl}) as expected theoretically by Eq. (\ref{tnl_dim}), $\tau_d(k)\sim k^{-1}$ (see $\S$ \ref{td}), whereas $\tau_l(k) \sim k^{-3/2}$ comes from the experimental dispersion relation. 
The time scale separation hypothesis of wave turbulence theory, $\tau_l(k) \ll \tau_{nl}(k) \ll \tau_d(k)$, is well satisfied experimentally for $k \gtrsim k_c=2\pi\times 10$ m$^{-1}$ as shown in Figure \ref{tnl1}(a) for two forcing amplitudes. For $k\lesssim k_c$, the separation scale breaks down since $\tau_l \approx \tau_{nl}  \ll \tau_d$. This scale corresponds to a wavelength $\lambda_c \simeq 10$ cm of the order of the forcing scale ($\gtrsim 10$ cm - see Fig. \ref{fig06}a), and thus is not involved in the cascade process. Finally, even if the nonlinear time scale is in good agreement with Eq.\ (\ref{tnl_dim}) and if the time scale separation is well satisfied experimentally within the inertial range, the scaling of the power-law spectrum with $k$ does not follow Eq. (\ref{DAtension}).

\section{Conclusion\label{conclusion}}
We have reported about experiments on nonlinear waves on the surface of a fluid covered by an elastic sheet. An optical method is used to obtain the full space-time wave field. It enables to reconstruct the dispersion relation of waves showing a transition between tension and bending waves. When the forcing is increased, a shift of the dispersion relation occurs. We show that this effect is due to an additional tension of the sheet induced by transverse nonlinear motion of a fundamental mode of the sheet. 

When the system is subjected to a random noise forcing at large scale, a regime of hydroelastic wave turbulence is observed. The spectrum of wave velocity then displays power laws of frequency and of wave number within an inertial range. This wave turbulence state corresponds theoretically to a direct cascade of energy flux through the scales. However, the frequency and wave number power law exponents of the spectrum are found in disagreement with the theoretical prediction. These exponents are found to be independent of the vessel size for the two tested configurations, of the strength of the forcing, and of the static sheet tension applied. To explain this discrepancy, several hypotheses of the theory have been experimentally tested to probe their validity domain in our experiment. Although an anisotropy of the wave field is induced by the forcing at large scale and weak forcing, it is not at the origin of this discrepancy. The time scale separation hypothesis has then been experimentally tested. The dissipation time of the waves, the nonlinear interaction time and the linear time of wave propagation have been measured at each scale. The separation of these time scales is well observed within the inertial range of the wave turbulence cascade. However, we have found that the dissipation occurs at all scales of the cascade contrary to the theoretical hypothesis. Although we can not discriminate the physical mechanism at the origin of the dissipation, the fact that it occurs at all scales could explain the discrepancy between the experimental and theoretical scalings of the spectrum. The occurrence of dissipation at all scales induces an ill-defined inertial range between the forcing and the dissipation and has been recently shown to be responsible for such a discrepancy in flexural wave turbulence on a metallic plate (\cite{Miquel2013,Humbert2013}). Possible non local interactions could be also put forward such as those involving the slow dynamics of a sheet fundamental mode that has been shown to modify the sheet tension at all scales. Moreover the possible interpretation of this mode dynamic as a condensation process, similar to what happen in 2D hydrodynamic turbulence (\cite{Sommeria}) or in optical wave turbulence (\cite{ResidoriReview}), is an open question and deserves further study. 

To conclude, a regime of wave turbulence has been observed in an original system even if some theoretical hypotheses are not fulfilled. We emphasize that dissipation occurring at all scales, and non local interactions should be included in future wave turbulence theories to have a more complete description of the experiments.

\begin{acknowledgments}
This work was supported by ANR Turbulon 12-BS04-0005. The authors thank M. Berhanu, M. Devaud, T. Hocquet, B. Miquel and M. Durand for fruitful discussions, and A. Lantheaume and C. Laroche for technical help.
\end{acknowledgments}

\bibliographystyle{jfm}

\bibliography{fgbib}

\end{document}